\documentclass[11pt]{article}
\pdfoutput=1
\usepackage{jheppub}
\usepackage{graphicx}
\usepackage{amssymb, amsmath, amssymb}
\usepackage{slashed}
\usepackage{hyperref}
\usepackage{caption}
\usepackage{xcolor}
\usepackage{dsfont}
\usepackage{verbatim}
\usepackage{subfig}
\usepackage{float,soul}
\usepackage{faktor}
\usepackage{extarrows}

\newcommand\ZZ{\mathbb{Z}}

\newcommand\Tr{\text{Tr}}

\newcommand\ra{\rightarrow}

\title{The Fractional Hall hierarchy from duality}

\author{Kristan Jensen$^a$ and}
\author{Amir Raz$^b$}

\affiliation{$^a$Department of Physics and Astronomy, University of Victoria, Victoria, BC V8W 3P6, Canada}
\affiliation{$^b$University of Texas, Austin, Physics Department, Austin TX 78712, USA}

\emailAdd{kristanj@uvic.ca}
\emailAdd{araz@utexas.edu}

\abstract{We show that a modified version of Son's Dirac composite fermion theory proposed by Seiberg et al gives a candidate unified description of the gapped and gapless fractional quantum Hall states within a single Landau level. Our main tool is the successive application of three-dimensional dualities to partially filled Landau levels of composite fermions, which imply that this theory has a complicated landscape of gapped vacua and critical points. This construction is the  Lagrangian, or effective field theory, analogue of the flux attachment procedure. The critical points exist at even denominator filling and are well-described by a Fermi surface for a weakly coupled composite fermion coupled to an abelian Chern-Simons theory. The gapped states include odd-denominator filling fraction states with an abelian Chern-Simons description which we show matches the one expected for hierarchy states, as well as non-abelian states at even-denominator filling that arise from pair instabilities of the composite fermion's Fermi surface.}

\begin{document}
\maketitle

\section{Introduction}

Ever since its first observation in \cite{firstFQH}, the fractional Hall effect has been an exemplar of strongly interacting physics. Many observed Hall states occur with filling fraction $\nu$ in the range $0<\nu <1$, and these states are expected to arise from interacting non-relativistic electrons in the lowest Landau level. Almost immediately after its discovery a successful and mature theory of the fractional Hall effect was quickly developed, starting with Laughlin's trial wavefunctions for gapped states of filling fraction $1/(2m+1)$ for $m=1,2,\hdots$~\cite{Laughlin_1983}. Using these trial wavefunctions Haldane~\cite{PhysRevLett.51.605} and Halperin~\cite{PhysRevLett.52.1583} separately constructed trial states for a vast ``hierarchy'' of gapped Hall states of fraction $p/q$. Subsequently, this approach matured into the composite fermion paradigm \cite{PhysRevLett.63.199,Moore:1991ks,PhysRevB.44.5246,HLR,jain2007composite}, whose original form describes a reorganized hierarchy~\cite{PhysRevLett.65.1502} of gapped states of odd-denominator filling as integer quantum Hall states of emergent ``composite fermions.'' The relation between these different composite fermions and the fundamental electrons, as well as with each other, is provided by a procedure known as ``flux attachment.'' All of these gapped states are described by abelian Chern-Simons theories~\cite{wen1992classification,Wen_Zee_92,Zee,tong2016lectures} that can be directly extracted from trial wavefunctions. The properties of these topological gaped states are well established, and many of them have been observed experimentally. For reviews of the fractional Hall effect see e.g. \cite{RevModPhys.71.S298,Hansson_QHrev,Zee,tong2016lectures}.

There are also observed gapless fractional Hall states, most notably at half filling. Halperin, Lee, and Read (HLR) proposed that near half filling fractional Hall physics is described by a field theory of non-relativistic composite fermions coupled to an emergent Abelian Chern-Simons gauge theory \cite{HLR}. This extremely successful effective field theory is able to predict both the gapped states in the Jain sequence \cite{PhysRevLett.63.199}, as well as a gapless Fermi liquid state at half filling. However, the HLR field theory lacks a fundamental symmetry of the quantum Hall system in the lowest Landau level: particle-hole symmetry. 

Both interacting electrons in the lowest Landau level and the collection of gapped states of the hierarchy preserve particle-hole symmetry,\footnote{Particle-hole symmetry maps a gapped state at filling $\nu$ to one at filling $1-\nu$.} so it would be natural for the gapless state at $\nu = 1/2$ to also preserve it. Inspired by this, Son proposed a particle-hole symmetric description near half filling of a composite Dirac fermion coupled to a dynamical gauge field \cite{Son_2015}. Son's field theory also predicts a gapless Fermi liquid state at half filling, just as HLR, though due to the fermion being a relativistic Dirac fermion the state is particle-hole symmetric.

In an initially unrelated development, there was substantial effort (see e.g.~\cite{Aharony:2011jz,Giombi:2011kc,Jain:2012qi,Aharony:2012ns,Jain:2013gza}) demonstrating the existence of dualities between Chern-Simons theories coupled to fundamental fermions, and Chern-Simons theories coupled to fundamental bosons, both in the `t Hooft limit. These ``$3d$ bosonization dualities'' were subjected to numerous quantitative tests. 
Aharony proposed a finite $N$ version~\cite{Aharony:2015mjs} of 3d bosonization, whose most extreme limit was the claim that a free Dirac fermion is dual to $U(1)_1$ Chern-Simons theory coupled to a charge-1 boson, together with a scalar potential tuned to a critical point. Assuming this equivalence the authors of~\cite{Seiberg_2016,Karch_2016} showed that it implies a whole web of dualities, including the known particle/vortex duality~\cite{Peskin:1977kp,Dasgupta:1981zz}. In particular, Seiberg, Senthil, Wang, and Witten (SSWW) showed that this basic bosonization duality implies one between a free fermion and a tweaked version of Son's proposal for a Dirac composite fermion, in which a Dirac fermion is coupled to two Chern-Simons gauge fields. They then put this model forward as a candidate description of the half-filled Landau level. 

Son's model and the SSWW theory are expected to describe the same gapless state at half filling. The two differ by certain global properties. For example, Son's theory has an additional $\mathbb{Z}_2$ one-form symmetry at all values of the filling fraction, including in its version of the fully empty and filled states where no such additional topological order should exist, while only the SSWW model obeys the spin/charge relation which we discuss in the next Section, which ensures that states of odd charge carry half-integer spin as appropriate for a model of microscopic electrons.

While this basic $3d$ bosonization duality remains unproven, there is evidence for it in the continuum summarized by Aharony~\cite{Aharony:2015mjs} and a strong argument for it from the lattice~\cite{Chen_2018,son2019duality}. There is a vast literature on other aspects of these finite $N$ dualities including e.g.~\cite{Hsin:2016blu,Aharony:2016jvv,Benini:2017dus,Komargodski:2017keh,Jensen:2017dso,Benini:2017aed,Jensen:2017bjo,Cordova:2017kue}.

The basic idea of this manuscript is to take the SSWW theory seriously as a candidate description of fractional Hall physics not just at half filling but throughout the lowest Landau level. We present significant evidence supporting this proposal.

We do this in steps. First, we argue that the fermion-fermion duality of~\cite{Seiberg_2016}, equating a free Dirac fermion with the SSWW theory, describes the desired system of interacting electrons in the lowest Landau level upon suitably deforming the free fermion side. The SSWW side is weakly coupled at half filling, where the dual ``composite fermion'' forms a Fermi liquid coupled to Chern-Simons gauge theory. This theory has perturbative vacua where the composite fermion forms Landau levels, some number of which are either fully filled or fully depleted. These gapped states are the basic Jain sequence, which we show is described by the abelian Chern-Simons theory expected from the point of view of trial wavefunctions. By partially filling a Landau level of the composite fermion, we find another strongly coupled description of interacting fermions in a Landau level. Using the fermion-fermion duality again we find another weakly coupled description of a dual composite fermion coupled to a more complex Chern-Simons gauge theory. This dualized description has a different gapless state, another Chern-Simons Fermi liquid, along with Jain states emanating from it when the dual fermion forms fully filled or depleted Landau levels.

Iterating this procedure we find a vast landscape of critical points at even-denominator filling fractions and gapped states at odd-denominator filling. The latter are described by unique abelian Chern-Simons theories. The former can generate certain non-abelian TQFTs when the Fermi liquid suffers BCS pairing instabilities. At filling $\frac{1}{2n}$ for $n=1,2,\hdots$ the gapless descriptions we find coincide with the proposal of Goldman and Fradkin for those filling fractions~\cite{Goldman_2018,Wang_2019,Nguyen_2021}.

This construction puts the composite fermion paradigm on a solid footing as a controlled quantum field theory description of the fractional Hall effect, weakly coupled in an appropriate duality frame, with the fermion-fermion duality of~\cite{Seiberg_2016} playing the role of the flux-attachment prescription. It also makes a variety of predictions, from the uniqueness of the topological order at odd denominator filling, the size of the composite Fermi surface at even denominator filling, and the parametric sizes of single-particle energy gaps.

We begin with a review in Section~\ref{S:review} about abelian Chern-Simons theory and fermions in 2+1-dimensions. We continue with an investigation of the half-filled gapless state of the SSWW theory and the Jain sequences emanating from it in Section~\ref{sec:jain}. By successive use of the fermion-fermion duality we recover a whole hierarchy of gapless and gapped states in Section~\ref{sec:main}, and study the states at even denominator in Section~\ref{sec:even}. We relegate some technical asides about one-form symmetry in the fractional Hall effect to the Appendix. 

\section{Background material}
\label{S:review}

We begin with some background information related to gauge fields and relativistic fermions in $2+1$ dimensions. Throughout we will follow the notation and naming conventions of \cite{Seiberg_2016}: background gauge fields will be denoted by uppercase Latin characters while dynamical gauge fields will be denoted by lower case ones. Background and dynamical spin$_c$ fields will be exclusively denoted by $A$ or $a$ respectively, while other Latin characters will be used for $U(1)$ gauge fields.

\subsection{Spin$_c$ Connections}

Systems described by interacting electrons obey what has been called the \emph{spin/charge relation}, whereby states of half-integer spin carry an odd electromagnetic charge, and states of integer spin carry even charge. A formal way of encoding this relation is the following. First, one takes the electromagnetic field to be a spin$_c$ connection~\cite{metlitski2015sdualityu1gaugetheory,seiberg2016gappedboundaryphasestopological,Seiberg_2016}, rather than a $U(1)$ gauge field. Locally a spin$_c$ connection is the same as a $U(1)$ gauge field, but globally it obeys 
\begin{equation} 
\label{eq:norm}
    \int_{C_2} \frac{dA}{2\pi} + \frac{1}{2}\int_{C_2} \omega_2 \in \ZZ\,,
\end{equation}
where $C_2$ is an oriented two-cycle and $\omega_2$ is the second Stieffel-Whitney class. As $\int_{C_2} \omega_2 \in \ZZ$, this means that $2A$ is a $U(1)$ gauge field, while $A$ may not obey the Dirac quantization condition. Second, at least for relativistic theories which enjoy general covariance, one mandates that the theory can be placed on any spin$_c$ manifold. The desired spin-charge relation automatically follows.

These two conditions imply that the basic Chern-Simons density for a spin$_c$ connection must be supplemented by a gravitational term to create a term such that $e^{iS}$ is well-defined on any spin$_c$ manifold,
\begin{equation}
   S = \int d^3x \,CS(A,g) \,, \qquad CS(A,g) = \frac{1}{4\pi}  AdA + \Omega(g)\,,
\end{equation}
where $\Omega(g)$ is a gravitational Chern-Simons term defined in \cite{seiberg2016gappedboundaryphasestopological,Hsin_2016}. Here $AdA$ is a shorthand for $\epsilon^{\mu\nu\rho}A_{\mu}\partial_{\nu}A_{\rho}$. The gravitational Chern-Simons term $\int d^3x \,\Omega(g)$ is also well-defined in its own right provided that its level is $8\mathbb{Z}$.

For $U(1)$ fields the standard Chern-Simons term of level $k$ describes a well-defined theory on any spin manifold, but it is only well-defined on a spin$_c$ manifold if $k$ is even. Mixed Chern-Simons terms between different  $U(1)$ fields, however, are allowed with arbitrary level. Finally, we can couple $U(1)$ and spin$_c$ fields by noticing that 
\begin{equation}
    CS(A+b,g) - CS(A,g) = \frac{1}{4\pi} b db + \frac{1}{2\pi} b  dA\,
\end{equation} 
describes a well-defined theory on any spin$_c$ manifold. However this coupling implies that the Chern-Simons level of $b$ must be the same as the integer coupling of $b$ to the spin$_c$ background modulo 2.

Overall, the most general abelian Chern-Simons theory described by a collection of dynamical $U(1)$ gauge fields $\{b_i\}_{i=1}^n$ coupled to a background spin$_c$ connection $A$ is given by
\begin{equation} \label{eq:CS}
    -\frac{K_{ij}}{4\pi}  b^i  d b^j + \frac{t_i}{2\pi} b^idA + k CS(A,g)+8 m \Omega(g)\,, \quad \quad K_{ij},t_i,k,m \in \ZZ\,, \quad K_{ii} - t_i \in 2\ZZ\,.
\end{equation}
The last condition, that  $K_{ii} - t_i$ (with no sum over $i$) is even, ensures the spin-charge relation.

We can see this in the following way. The quasiparticles of this theory are described by the line operators 
\begin{equation}
    \mathcal{W}_{C_1, \xi } = e^{i \int_{C_1} \xi_i b^i }\,, \qquad \xi \in \ZZ^n\,,
\end{equation}
where $C_1$ is a 1-cycle. These operators have braiding, spin (in the sense of self-statistics) and charge
\begin{equation}
    B(\xi,\zeta) = e^{2\pi i \xi K^{-1} \zeta}\,, \qquad S(\xi) = \frac{1}{2}\xi^T K^{-1} \xi\,, \qquad Q(\xi) =  \xi^T K^{-1} t\,.
\end{equation}
Quasiparticles with non-trivial braiding are anyons, while the transparent operators, those with trivial braiding, correspond to the worldlines of particles constructed from the microscopic electrons. The spin/charge relation mandates that such transparent operators have $Q-2S$ even. The transparent operators are those with $\xi = K \zeta$ for $\zeta \in \ZZ^n$, for which $Q - 2S =\zeta^T (K\zeta - t)$ which is indeed even on account of $K_{ii}-t_i$ being even.

We will also consider theories with dynamical spin$_c$ connections alongside dynamical $U(1)$ gauge fields. To map these back to more traditional Abelian Chern-Simons theories we will use the fact that on a manifold without boundary there is a Hubbard-Stratonovich transformation that allows us to replace a level-1 Chern-Simons term for a spin$_c$ connection with a dynamical $U(1)$ gauge field at level 1,\footnote{This equivalence was worked out in detail in Appendix B of \cite{Seiberg_2016}.}
\begin{equation}
   CS(a,g) \qquad \cong  \qquad -\frac{1}{4\pi} b  db + \frac{1}{2\pi}  b da \,.
\end{equation}
As a corollary, we can replace a level-$n$ Chern-Simons term for a spin$_c$ connection with an equivalent Chern-Simons description with $|n|$ $U(1)_1$ fields $b_i$:
\begin{equation} \label{eq:CSAtoU1}
    nCS(a,g) \quad \cong  \quad
    -\text{sgn}(n)\sum_{i=1}^{|n|} \frac{1}{4\pi}b_i  db_i + \frac{1}{2\pi} \left(\sum_{i=1}^{|n|}b_i \right) da\, .
\end{equation} 

While we can always map dynamical spin$_c$ connections to a purely $U(1)$ Chern-Simons theory, it will be useful to understand spin$_c$ Chern-Simons theories in their own right. Dynamical spin$_c$ connections are locally the same as $U(1)$ gauge fields, so much of the intuition and results for Abelian Chern-Simons theories applies. For example Chern-Simons theories have a global 1-form symmetry consisting of flat shifts to the gauge fields that are a symmetry of the action but are not large gauge transformations. These same transformations will still be the 1-form symmetry of the action when some of the fields are spin$_c$, rather than $U(1)$, as they do not change the normalization condition. For a detailed discussion see Appendix~\ref{A:oneForm}.

The main difference between spin$_c$ and $U(1)$ connections are the allowed line operators and their quantum numbers. For an ordinary $U(1)$ field the normalization $\int_{C_2} db \in 2\pi \ZZ$ implies that $\int_{C_1} b$ is only well defined modulo $2\pi \ZZ$ by the fundamental theorem of calculus. Thus for a spin$_c$ connection $a$, the well defined Wilson line operators are modified by the normalization \eqref{eq:norm} to \cite{Seiberg_2016}
\begin{equation}
    \mathcal{W}_{C_1, n,q } = e^{i \int_{C_1} \left( n a + \frac{q}{2} \omega \right) }, \qquad n,q \in \ZZ, 
\end{equation}
where $\omega$ is an abelian spin connection constructed from the worldline and the spin connection of the space, and $n = q \mod 2$. This modifies the spin (and hence braiding) of these lines by $q/2 = n/2 \mod \ZZ$.  

\subsection{Fermion determinants}
\label{sec:fermdet}

In this paper we frequently consider a Dirac fermion $\psi$ coupled to a background spin$_c$ connection $A$
\begin{equation}
    \mathcal{L} = i \bar{\psi} \slashed{D}_A \psi = i \bar{\psi} \gamma^\mu\left(\partial_\mu -i A_\mu\right) \psi\,.
\end{equation}
We can integrate out the fermion leading to the partition function
\begin{equation}
    Z_\psi (A) = \det(\slashed{D}_A)\,,
\end{equation}
however this determinant needs to be regularized. The regularization of \cite{Witten_2016,Seiberg_2016} leads to 
\begin{equation} \label{eq:ZPsi}
    Z_\psi (A) = |Z_\psi(A)| e^{-i\frac{\pi \eta(A)}{2}}\,,
\end{equation}
where $\eta(A)$ is the Atiyah-Patodi-Singer $\eta$-invariant, which counts the difference between the number of positive and negative eigenvalues of $\slashed{D}_A$ in a regularized way.\footnote{The original definition of $\eta$ was $\eta = \lim_{s\ra 0} \sum_k \text{sgn}(\lambda_k)|\lambda_k|^{-s}$, but other regularization are equivalent \cite{atiyah1975spectral,Witten_2016}.} 

Adding a Dirac mass term generates an additional phase of $e^{i \text{sgn}(m) \frac{\pi \eta(A)}{2}}$. Overall in these conventions we get that $Z_\psi(A,m\to\infty) = 1$ while $Z_\psi(A,m\to-\infty) =  e^{-i \pi \eta(A)} = e^{-i\int CS(A,g)}$ by the Atiyah-Patodi-Singer index theorem. Note that while $e^{-i \pi \eta(A)}$ describes a Chern-Simons effective action $e^{-i \pi \eta(A)/2}$ has no such interpretation since $e^{-i \int CS(A,g)/2}$ is ill-defined.

Choosing a different regularization amounts to a choice of different local counterterms, which for this theory are integer multiples of $CS(A,g)$. The gapless theory has a time reversal symmetry. However, the fact that the partition function \eqref{eq:ZPsi} is not real, and cannot be made real by adding local counterterms, implies that the time reversal symmetry is anomalous. The time reversal symmetry of 2+1-dimensional massless Dirac fermions is equivalently a parity symmetry, and this is the famous ``parity anomaly''~\cite{Niemi:1983rq,Redlich:1983kn,Alvarez-Gaume:1984zst}, which is really a mixed anomaly between time-reversal/parity and the $U(1)$ symmetry. This anomaly implies that there is no way to define the massless fermion in a way that preserves both time-reversal and $U(1)$. Here we are keeping $U(1)$ sacrosanct at the expense of time-reversal.

One way to cure this anomaly is to consider the system as a boundary of a $3+1$ dimensional bulk \cite{seiberg2016gappedboundaryphasestopological,Seiberg_2016}. Then we can extend $A$ and the metric to the bulk and add a bulk action for $A$
\begin{equation}
    S_\text{bulk} = \frac{1}{8\pi}\int dA \wedge dA + \frac{1}{192 \pi}\int \Tr~ R \wedge R\,,
\end{equation}
where $R$ is the Riemann curvature two-form. The variation of this term on the $2+1$ dimensional boundary cancels the phase in \eqref{eq:ZPsi}, and so make the theory time reversal invariant. 

\subsection{Landau levels of a Dirac fermion and relativistic Wen-Zee terms}

As our interest lies in quantum Hall states we will turn on a uniform background magnetic field encoded in the background spin$_c$ connection $A$. In an external field the fermions form Landau levels, and on a spatial torus the energy spectrum is
\begin{equation} \label{eq:spect}
    E_n = \text{sgn}(n+ \epsilon m)\sqrt{m^2 + 2B|n|}\,, \qquad n\in \ZZ\,,
\end{equation}
where $n$ denotes the landau level, $m$ is the mass of the fermion, $B$ is the external magnetic field, and $1\gg \epsilon>0$ is used to set $\text{sgn}(n=0) = \text{sgn}(m)$. For most of the paper we take the fermion to be massless, in which case the energy levels are $ E_n = \text{sgn}(n)\sqrt{2B|n|}$. Each of these Landau levels has a degeneracy given by the number of units of magnetic flux $N_\phi = \frac{B V}{2\pi}$ where $V$ is the spatial volume.

An interesting question is what happens to the fermion partition function when we tune the chemical potential so that the system lies in a gap of the spectrum. Taking the gap to be large by increasing the external field only the phase of the partition function remains. This phase can be evaluated using spectral flow as it is proportional to the regularized difference between positive and negative eigenvalues \cite{Witten_2016}. 

To find the phase obtained by filling a single Landau level, we can simply add a Dirac mass term. For negative mass we know the partition function is $Z_\psi(A,m\to -\infty) =  e^{-i\int d^3x\,CS(A)}$, and based on the spectrum \eqref{eq:spect} all Landau levels with $n\leq 0$ will be full. For positive mass, on the other hand, the partition function is $Z_\psi(A,m\to \infty) = 1$, and Landau levels with $n< 0$ will be full. Thus filling a single Landau level gives a factor of $e^{-i\int d^3x \,CS(A)}$. On a torus all of the Landau levels have identical degeneracy. Thus, if instead we fill all Landau levels with $n< N$ then the phase will be $e^{-iN\int d^3x\,CS(A)}$. This is simply the integer quantum Hall effect: the filling is quantized to be the number of filled Landau levels and the low energy effective action for these gapped states is a level $N$ Chern-Simons term for the background field $A$.

There is, however, a correction to the degeneracy of each Landau level when the constant time slice is a curved surface, called the shift \cite{Wen_Zee_92}. For example on the round sphere the $n$th Landau level has a degeneracy $N_\phi+2|n|$ (see e.g.\cite{Hasebe_2016}), which leads to a constant shift in the charge of the filled level. In non-relativistic Hall physics the analogous offset implies the existence of the Wen-Zee term in the effective action of the gapped state $ \frac{s}{2\pi} \int A \wedge d \tilde{\omega}$, where $\tilde{\omega}$ is the spatial spin connection \cite{Wen_Zee_92} (discussed below). In this case the offset implies a relativistic version of the Wen-Zee term which can be constructed as follows. The magnetic field defines a nowhere-vanishing timelike unit vector field $u$ which locally points in the direction $\star dA$. Coupling the microscopic fermion to a slowly varying curved space described by a dreibein and spin connection, we can then orient the time component of the dreibein with $u$, and use it to contract the $SO(1,2)$ spin connection $\omega$ to a $U(1)$ spin connection $\omega' = -\frac{1}{2}\epsilon^{abc} \omega_{ab} u_c$, which transforms under local rotations that preserve the magnetic field as $\omega' \to \omega' + dv$. The offset in the Landau level degeneracy implies a shift in the charge $\Delta Q = |n|\chi$ when the 3$d$ space is of the form $\mathbb{R}\times \Sigma$ where the compact spatial manifold $\Sigma$ is a surface of characteristic $\chi$ through which we thread a large magnetic field. This in turn implies a relativistic version of the Wen-Zee term in the effective action of the filled level, $ \frac{|n|}{2\pi} \int A \wedge d\omega'$. Note that this term can only be defined due to the magnetic field breaking Lorentz invariance. A similar construction appeared in \cite{Golkar_2014,Golkar_2015}, and one can verify that their Euler current is equivalent to $\star d\omega'$ in our formulation using Cartan's structure equations. 

We treat $\omega'$ as a properly normalized $U(1)$ connection. This is consistent with the fact that on a product spacetime $\mathbb{R}\times \Sigma$ threaded by constant flux we have $\int_{\Sigma} \frac{d\omega'}{2\pi} = \chi_{\Sigma}$, the Euler characteristic of $\Sigma$. Now recall that on a spin$_c$ manifold $\frac{1}{2\pi} A db$ is not globally well-defined, but $\frac{1}{2\pi} A db + \frac{1}{4\pi} bdb$ is. It is unclear to us if this statement is relaxed when considering the subset of all spin$_c$ manifolds that admit a globally defined nowhere vanishing vector field. Assuming that it holds, and viewing $\omega'$ as a $U(1)$ connection, we infer that the relativistic Wen-Zee term must be accompanied by a Chern-Simons term for $\omega'$ to be well-defined,
\begin{equation}
	\frac{1}{2\pi}A d\omega' + \frac{1}{4\pi} \omega' d\omega'\,,
\end{equation}
and that integer multiplies of $\frac{2}{4\pi}\omega' d\omega'$ are also well-defined. 

Independent of these considerations, we can directly calculate the effective action of a filled or depleted Landau level. We can already infer the coefficients of $CS(A)$ and $A d\omega'$ in that effective action by spectral flow, using the degeneracy $N_{\phi} + 2|n|$ of the $n$th Landau level on the sphere. We can also calculate the coefficient of $\omega' d\omega'$ by turning on a large Dirac mass, taking a non-relativistic limit, and matching to the action of filled non-relativistic Landau levels obtained in~\cite{Abanov_2014}. Those authors considered a non-relativistic a spinless fermion coupled to a background $U(1)$ field $B_{\mu}$ and in a weakly curved spacetime described by
\begin{equation}
	\label{E:NRfermion} 
	S_{\rm NR} = \int d^3x \sqrt{\gamma}\left( i v^{\mu} \psi^{\dagger}(\partial_{\mu} - i B_{\mu}) \psi - \frac{h^{\mu\nu}}{2m} (\partial_{\mu} + i B_{\mu})\psi^{\dagger} (\partial_{\nu}-i B_{\nu}) \psi\right)\,,
\end{equation}
where $v^{\mu} $ is a nowhere-vanishing vector field, $h^{\mu\nu}$ is a rank-2 symmetric positive-semidefinite inverse spatial metric, and $\gamma^{\mu\nu} = v^{\mu}v^{\nu} + h^{\mu\nu}$. From the background fields one can define an abelian spin connection $\tilde{\omega}_{\mu}$. To quadratic order in fluctuations around a flat space integer Hall state those authors obtained the effective Lagrangian of the gapped state whose Chern-Simons like terms are
\begin{align}
\begin{split}
	L_{\rm NR,eff} = \sum_{k=1}^n CS\left( B + \left(k-\frac{1}{2}\right)\tilde{\omega},\gamma\right) \,,
\end{split}
\end{align}
where $k$ indexes the $k$th filled Landau level. This result is related to the relativistic one in the following way. We start with a Dirac fermion in a weakly curved background spacetime and coupled to a spin$_c$ field $A_{\mu}$, turn on a large mass $m$, a background magnetic field $B$, and tune the chemical potential to zoom in on a Landau level. Half of the fermion components become very heavy, which upon integrating it out leaves a non-relativistic spinless fermion behind with an effective action of the form~\eqref{E:NRfermion} with the identification $B_{\mu} = A_{\mu} - \frac{1}{2}\omega'_{\mu}$ along with $\tilde{\omega}_{\mu} = \omega'_{\mu}$. 

In our convention for the $\eta$ invariant, the low-energy effective Lagrangian of the gapped state between the $n-1$th and $n$th Landau levels is then 
\begin{align} \label{eq:intQH}
    L_{n} &=    \text{sgn}(n)\sum_{j=1}^{|n|}\,CS\left(A + \left( j - \frac{\text{sgn}(n)+1}{2}\right)\omega',g\right) \\
    \nonumber
    &=   n CS(A,g)+\frac{|n|(n-1)}{4\pi}  A d\omega' + \frac{n(n-1)(2n-1)}{24\pi} \omega'd \omega'\,,
\end{align}
with the term $j$ corresponding to the $j-1$th filled Landau level for $n>0$ and $-j$th depleted level for $n<0$. This is consistent with the global considerations we described above.

\begin{figure}
    \centering
    \includegraphics[width=0.7\linewidth]{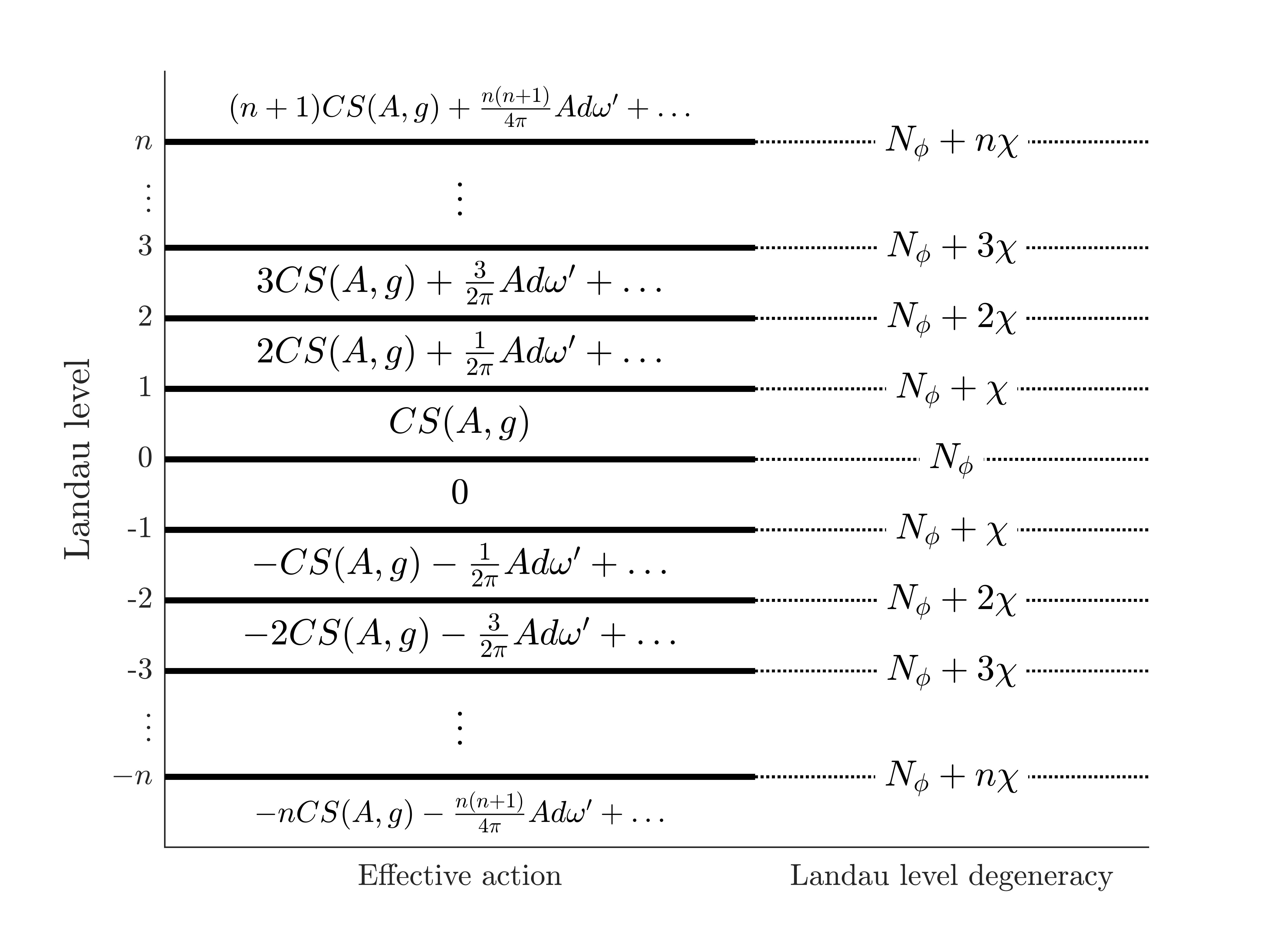}
    \caption{The effective action in the gaps between Landau levels in our convention, as well as the degeneracy of each Landau level which is universal. The change in the effective action as one crosses a Landau level is universal, and completely determined by the degeneracy. The ellipsis indicate the purely gravitational $\omega' d\omega'$ terms.}
    \label{fig:intqh}
\end{figure}

We sum up our convention for the effective action in the gaps, along with the Landau level degeneracy in Fig.~ \ref{fig:intqh}. Other conventions differ from ours by local counterterms, which amounts to changing the level of the Chern-Simons theory and the Wen-Zee term by some integers. However, the difference in the EFT as one fills an additional level is completely determined by the degeneracy of the Landau level crossed, and cannot be altered by local counterterms.

\section{The basic fermion-fermion duality and the Jain sequence}
\label{sec:jain}

Inspired by 3d bosonization and level-rank dualities, Seiberg, Senthil, Wang and Witten (SSWW) conjectured a fermion-fermion duality in \cite{Seiberg_2016}:\footnote{Note that we have chosen to apply a charge conjugation operation to the right hand side relative to the version listed in \cite{Seiberg_2016} to make some of the computations simpler. }
\begin{equation}
    \label{eq:duality}
    i\bar{\psi}\slashed{D}_A\psi + CS(A,g) \qquad \Longleftrightarrow \qquad i\bar{\chi}\slashed{D}_a\chi - \frac{1}{2\pi} a db - \frac{2}{4\pi} b db + \frac{1}{2\pi} b dA \,.
\end{equation}
Implicitly the right-hand-side includes irrelevant interactions, like Maxwell terms for $a$ and $b$, but the resulting theory flows to a critical point in the deep IR and this point is secretly a free fermion. 

Both sides of this duality have a charge conjugation symmetry as well as anomalous time-reversal and parity symmetries. Time-reversal forbids a mass term for $\chi$. We can define genuinely time-reversal and parity-invariant theories by coupling both sides of the duality~\eqref{eq:duality} to a $3+1d$ bulk, resulting in
\begin{equation}
    \label{eq:dualityT}
    i\bar{\psi}\slashed{D}_A\psi + \frac{1}{2} CS(A,g) \qquad \Longleftrightarrow \qquad i\bar{\chi}\slashed{D}_a\chi - \frac{1}{2\pi} a db - \frac{2}{4\pi} b  db + \frac{1}{2\pi} b dA - \frac{1}{2} CS(A,g) \,,
\end{equation}
although we will not use this form in what follows.

The left hand side of the duality~\eqref{eq:duality} is a free fermion coupled to a background spin$_c$ field $A$, while the right hand side is an interacting theory of a fermion $\chi$ coupled to a dynamical spin$_c$ connection $a$ and a dynamical $U(1)$ gauge field $b$. SSWW conducted many tests of this proposal including the matching of symmetries and anomalies, the quantum numbers of various local operators, and the phase diagram at large fermion mass.

The right hand side has no small parameter and so this duality is colloquially understood as the statement that at $A=0$ the right side is strongly interacting, but is free in the dual variable $\psi$ in the deep infrared. As noted by Son~\cite{Son_2015}, the free fermion side is in fact related to the problem we wish to study, that of interacting non-relativistic electrons in a strong applied field, with $\psi$ playing the role of the microscopic electron. That relation arises in the following way. First, let us consider the duality on a large spatial torus of volume $V$ and turn on a large background magnetic field for $A$ with $N_{\phi} \gg 1$ units of flux, so that the spectrum on the left side becomes that of highly degenerate Landau levels with an interlevel spacing $\sim \sqrt{B}$. At low energies only the lowest Landau level survives, and within a level there is no distinction whether the fermion is relativistic or not. Now turn on interactions, say a 3+1-dimensional Coulomb interaction between the $\psi$'s, that are weak compared to the external field, so that the interactions do not mix the states in the different Landau levels. Within that approximation the problem we land on is precisely that of interacting non-relativistic electrons in the lowest Landau level.

Let us briefly pause to discuss symmetries. Before adding interactions the non-relativistic lowest Landau level has an anti-unitary particle hole symmetry which sends states of filling fraction $\nu$ to those of filling $1 - \nu$. We have in mind interactions, like the Coulomb interaction, that respect this property, so that the interacting theory has a genuine symmetry at half-filling. Realizing this theory starting from the duality~\eqref{eq:duality}, the applied field breaks the anomalous time-reversal and parity symmetries of the left-hand-side, but leaves intact the anomalous combination $\mathsf{CT}$. It is this symmetry that becomes the particle-hole symmetry of the resulting non-relativistic problem, and a genuine symmetry of the half-filled state.

After these deformations the left side of the duality now becomes strongly interacting since the only scale available is set by the interactions. What happens to the right side? We will call the dual fermion $\chi$ the ``composite'' fermion. There is now a background magnetic field for $A$ and the interactions on the left side can in principle be mapped to an unknown deformation $\mathcal{O}$ of the right side. Naively we cannot proceed without knowing more about the precise operator $\mathcal{O}$ generated, but surprisingly we do not need to know much about $\mathcal{O}$ at half filling, i.e. when the charge is $Q = \frac{1}{2}N_{\phi}$. Temporarily ignoring the deformation $\mathcal{O}$, at this value the dual fermion $\chi$ forms a large, gapless Fermi surface and so the right side of the duality becomes weakly interacting. 

To see this note that the charge current on the right side of the duality is $J^{\mu} = \frac{1}{2\pi} \epsilon^{\mu\nu\rho}\partial_{\nu}b_{\rho}$. Moreover the time component of $b_{\mu}$ acts as a Lagrange multiplier enforcing the condition
\begin{equation}
	\epsilon^{ij} \partial_i a_j + 2 \epsilon^{ij} \partial_i b_j - \epsilon^{ij} \partial_i A_j = 0\,.
\end{equation}
Putting these two statements together at half filling we see that the spatial fluxes of $a$ and $b$ are constrained to be
\begin{equation}
	\oint \frac{db}{2\pi} = \frac{N_{\phi}}{2}\,, \qquad \oint \frac{da}{2\pi} = 0\,.
\end{equation}
The composite fermion therefore feels no average magnetic field. There will be local quantum fluctuations of these fluxes, but they must spatially average to zero. Moreover, the time component of $a_{\mu}$ acts as a Lagrange multiplier enforcing that the composite fermion charge coincides with the electric charge, $Q_{\chi} = \oint \frac{db}{2\pi}$, and therefore the total composite fermion charge is large, $Q_{\chi} = \frac{N_{\phi}}{2}\gg 1$. The Chern-Simons terms effectively gap the statistical fields $a$ and $b$, and as a result we expect the right side to flow to a Fermi liquid, coupled to Chern-Simons gauge fields, which is weakly coupled thanks to the large size of the Fermi surface whose radius is set by the inverse magnetic length. The resulting theory is expected to have the same critical exponents as the Dirac composite Fermi surface presented by Son in~\cite{Son_2015} with a similar but non-particle-hole symmetric Fermi liquid arising in the HLR theory~\cite{HLR}.

In presenting this argument we have ignored the deformation $\mathcal{O}$. Now we restore it. By assumption we consider interactions which preserve particle-hole symmetry, here $\mathsf{CT}$, and so $\mathcal{O}$ must be $\mathsf{CT}$-invariant. Usually $\mathcal{O}$ will then be a $\mathsf{CT}$-invariant irrelevant deformation of the resulting Chern-Simons Fermi liquid, by the usual results for the robustness of the Landau Fermi liquid effective theory against perturbations~\cite{Polchinski:1992ed,Shankar:1993pf}. For example it might include Maxwell terms for the statistical fields. However this need not always be the case. If $\mathcal{O}$ includes a BCS pairing instability then it will drive the formation of a gap so that the right hand side flows to a massive phase, a certain topological superconductor.

Half filling is not the only filling fraction where the composite fermion is weakly coupled. Interactions are also weak when the composite fermion is gapped. This happens in those states when $\chi$ feels a large magnetic field, forms Landau levels, and where a number of those levels are completely filled. As argued in Section \ref{sec:fermdet}, when a Dirac fermion completely fills a Landau level the fermion can be integrated out and replaced with a Chern-Simons term whose level depends on the number of filled Landau levels. This is valid at energy scales much smaller than the gap between Landau levels, which is of the order of the square-root of the effective magnetic field for $a$ by~\eqref{eq:spect}. The state where $n-1$ Landau levels are completely filled corresponds to an average charge density $\rho_\chi = \left( n + \frac{1}{2}\right)\frac{B_a}{2\pi}$, where $B_a$ is the average magnetic field for $a$. In our convention for $n$ entirely filling the lowest Landau level for $\chi$ corresponds to $n=1$. Using similar arguments to what we did above at half filling, we find
\begin{equation}
     Q_\chi - \oint \frac{da}{2\pi} = \oint\frac{db}{2\pi}\, , \qquad 2\oint \frac{db}{2\pi} + \oint\frac{da}{2\pi} = N_{\phi}, \qquad \oint \frac{db}{2\pi} = \nu N_{\phi}\,,
\end{equation}
where $\nu$ is the filling fraction of the state. Solving these equations tells us that these states have the filling fractions of the Jain sequence, 
\begin{equation}
    \nu = \frac{n}{2n+1}\,, \qquad \qquad \oint \frac{da}{2\pi} = \frac{N_{\phi}}{2n+1}\,.
\end{equation}
Note that since $n$ can be positive or negative that these results simultaneously describe the Jain sequences with $\nu = \frac{n}{2n\pm 1}$ with $n=0,1,2,..$. Moreover the states with negative $n$ have a magnetic field $B_a$ oppositely aligned to the physical one. In the non-relativistic composite fermion picture the sequence with positive $n$ arises from attaching flux to electrons, while the sequence with negative $n$ comes after attaching flux to holes. Here both sequences arise from one and the same Dirac composite fermion. To go from one step in the sequence to the next we completely deplete or completely fill one more of its Landau levels.

Note also that the average magnetic field for $a$ becomes weaker at large $|n|$. As we assumed the interaction scale is weak relative to the external magnetic field, the interaction strength is, at least for modest $|n|$, also much smaller than the spectral gap which is proportional to $\sqrt{B_a}$. So we expect the interactions to not mix these composite fermion Landau levels for low enough $|n|$, and thus to not alter the low energy TQFT description. However this will not be true at large $|n|$ where the interaction scale is comparable to the interlevel spacing. Without detailed knowledge of the deformation $\mathcal{O}$ we cannot ascertain the crossover between these two behaviors. The existence of such a crossover is also a prediction of the standard non-relativistic CF framework~\cite{PhysRevLett.63.199}.

The explicit Chern-Simons description of this phase is
\begin{equation} \label{eq:CS_Jain}
     n CS(a,g) - \frac{1}{2\pi}  a  db - \frac{2}{4\pi}b  db + \frac{1}{2\pi} b  dA \,.
\end{equation}
Note that $a$ is a spin$_c$ field while $b$ is a $U(1)$ field, and so the term $CS(a,g)$ also contains gravitational terms. This theory looks like a coupled Chern-Simons theory with $K$ matrix $K = \begin{pmatrix} 2 & 1 \\ 1 & -\mathbf{n}\end{pmatrix}$, and a charge vector $t = \begin{pmatrix} 1 & 0 \end{pmatrix}^T$. If both $a$ and $b$ were $U(1)$ fields then the ensuing Chern-Simons theory would violate the spin-charge relation leading to incorrect predictions for the spectrum of massive quasiparticles. However in fact this theory obeys the spin-charge relation because $a$ is a spin$_c$ connection. To emphasize this point we bold the diagonal entries in the $K$ matrix corresponding to spin$_c$ fields.

While the TQFT in \eqref{eq:CS_Jain} seems different than the standard Chern-Simons description of the Jain sequence or the equivalent hierarchy construction \cite{Wen_1995}, they are in fact the same on a closed space as we will show shortly. Before doing so we would like to emphasize the strength of the duality \eqref{eq:dualityT}: from it we can completely derive the integer quantum Hall states, the effective theory of the Jain sequence, and the appearance of a Fermi liquid at half filling. We summarize all these results in Table \ref{tab:dual}.

\begin{table}[]
    \centering
    \begin{tabular}{c|c|c|c|c}
    & \multicolumn{2}{|c|}{Interaction strength} & &\\ 
        Filling & Left side & Right Side & Gap? & Phase \\
         \hline
        $\nu = 0,\pm1, \pm2 ,\ldots$ & Weak & Strong & Gapped & $\nu CS(A,g)$ \\
        $\nu = 1/2$ & Strong & Weak& Gapless& Fermi liquid\\
        $\nu = \frac{n}{2n+1}, \quad n \in \ZZ$ & Strong & Weak & Gapped & Coupled Abelian CS theory
    \end{tabular}
    \caption{The different quantum Hall phases described by the duality \eqref{eq:dualityT}. The integer Hall states are weakly coupled in the original free fermion picture, while the Fermi liquid at half filling and the Jain sequence are weakly coupled in the dual composite fermion description.}
    \label{tab:dual}
\end{table}

\subsection{Matching to the expected TQFT}
\label{sec:Jain_K}

We presently match the action \eqref{eq:CS_Jain} to the standard Chern-Simons theory of the Jain sequence. A similar matching for abelian bosonic fractional Hall states was performed in~\cite{Geraedts_2017}. This in turn is equivalent to the TQFT description of single-level hierarchy states with the filling fraction $\frac{n}{2n+1}$~\cite{PhysRevLett.65.1502,Wen_1995}. Our argument goes through a space without boundary. The standard TQFT description of the Jain sequence at filling $\nu = \frac{n}{2n+1}$ is given by a coupled $U(1)$ Chern-Simons theory \eqref{eq:CS} of $|n|$ $U(1)$ gauge fields with the following $K$ matrix, $t$ vector, and spin vector,
\begin{equation}
 \label{eq:JainOG}
    K= \text{sgn}(n)I_{|n|} + 2 J_{|n|} \,,  
    \quad 
    t = \begin{pmatrix}
        1 & 1 & 1 &  \cdots
    \end{pmatrix}^{\rm T} = 1_{|n|}^{\rm T}\,,
    \quad 
    s = \begin{cases}  \begin{pmatrix} 3/2 & 5/2 & 7/2 & \cdots \end{pmatrix}^{\rm T}  \,, & n>0\,, \\ -\begin{pmatrix} 1/2  & 3/2 & 5/2 & \cdots \end{pmatrix}^{\rm T}\,, & n<0\,, \end{cases}
\end{equation}
with $I_{|n|}$ the $|n| \times |n|$ identity matrix, and $J_{|n|}$ the $|n| \times |n|$ matrix with all entries being $1$. The corresponding Lagrangian is
\begin{equation}
	-\sum_{i,j=1}^{|n|} \frac{1}{4\pi}K_{ij}  c_i dc_j + \sum_{i=1}^{|n|}\frac{1}{2\pi}( t_i c_i dA + s_i c_i  d\omega')\,.
\end{equation}

To match this form we can use \eqref{eq:CSAtoU1} to replace the level $n$ Chern-Simons term for $a$ in \eqref{eq:CS_Jain} with $n$ $U(1)$ fields $c_i$, resulting in
\begin{equation} \label{eq:K_Jain}
     -\sum_{i=1}^{|n|} \frac{\text{sgn}(n)}{4\pi} c_i dc_i + \frac{1}{2\pi}  \left(\sum_{i=1}^{|n|}c_i \right)  da - \frac{1}{2\pi}  a db - \frac{2}{4\pi} b db + \frac{1}{2\pi} b  dA \,.
\end{equation}
Here it was crucial that we work on a space without boundary. With a boundary perhaps there are boundary conditions on the various fields so that a similar path integral identity exists without requiring the introduction of additional edge modes. Now we notice that $a$ appears linearly in the action, and so is a Lagrange multiplier which can be integrated out enforcing the constraint $b = \sum_i c_i  + (\text{gauge})$. This leads to
\begin{equation} 
\label{E:matchingJain}
     -\sum_{i,j=1}^{|n|} \frac{1}{4\pi}(\text{sgn}(n)\delta_{ij}+2) c_i  dc_j + \frac{1}{2\pi} \sum_{i=1}^{|n|} c_i  dA \,.
\end{equation}
This is the standard TQFT description~\eqref{eq:K_Jain} of the Jain sequence as claimed.

We can also match the spin vector, and hence shift, for these states to the expected value in~\eqref{eq:K_Jain} for the Jain sequence. In Section \ref{sec:fermdet} we showed that the $n$th filled or depleted Landau level of a Dirac fermion coupled to an external field $A$ is described by the term $\frac{|n|}{2\pi} \int A \wedge d\omega'$ in its low-energy effective action. In the present instance this implies the presence of a Wen-Zee like term in the TQFT description of $(a,b,c_i)$,
\begin{align}
\begin{split}
     -\sum_{i=1}^{|n|} \frac{\text{sgn}(n)}{4\pi}c_i  dc_i & + \frac{1}{2\pi} \left(\sum_{i=1}^{|n|}c_i \right)  da + \sum_{j = 1}^{|n|} \frac{\text{sgn}(n)}{2\pi}\left(j + \frac{\text{sgn}(n)-1}{2}\right)  c_j  d \omega' 
    \\
    & - \frac{1}{2\pi} a  db - \frac{2}{4\pi}b  db + \frac{1}{2\pi}  b  dA \,.
\end{split}
\end{align}
Integrating out $a$ again enforces $b = \sum_{i=1}^{|n|}c_i + (\text{gauge})$, leading to a residual theory of the $c_i$. The only difference with what we wrote in~\eqref{E:matchingJain} is the Wen-Zee term
\begin{equation}
	\sum_{i=1}^{|n|}\frac{\text{sgn}(n)}{2\pi}\left( i+ \frac{\text{sgn}(n)-1}{2}\right) c_i d\omega'\,.
\end{equation}
This gives a spin vector of $s_j =  j$ for $n>0$ and $s_j = -(j-1)$ for $n<0$ , which do not match their expected values in~\eqref{eq:K_Jain}. However, recall that $A$ in our analysis is a spin$_c$ field, rather than the standard $U(1)$ background typically considered in the condensed matter literature. Given a particular spin structure and spin connection on an orientable manifold, we can take $A$ to be $A = B + \frac{\omega'}{2}$, where $B$ is the background electromagnetic field and $\omega'$ is the spatial spin connection. Rewriting our TQFT this way leaves the $K$-matrix and charge vector (defined through the coupling of the statistical fields to $B$) but redefines the spin vector as $s\to s + \frac{t}{2}$. With this identification we match the expected spin vector and thus also the shift on the Jain sequence.

This being said, from the point of view of the low-energy Chern-Simons theory it seems that the spin vector is only defined up to identification $s\to s+t$. The reason for this is that we can redefine the background field $A$ as $A\to A+p \omega'$ for any integer $p$, which shifts the spin vector as $s\to s+p t$. This identification acts on the shift as $\mathcal{S} \to \mathcal{S}+2$.

We can also map the line operators directly from the TQFT description in terms of $(a,b)$ to those of the standard description. The line operators in the spin$_c$ description are characterized by integers $n_b$ and $n_a$ and take the form
\begin{equation}
    W_{n_b,n_a} = \exp \left( i \int\left(n_b b + n_a a  \right)\right)\,.
\end{equation}
They have self-statistics and charge modulo 1
\begin{equation}
\label{E:basicJainLines}
    S(n_b,n_a) = \frac{n n_b^2 + 2n_b n_a - 2n_a^2}{2(2n+1)} + \frac{n_a}{2}, \qquad \qquad Q(n_b,n_a) = \frac{n n_b + n_a}{2n+1}\,.
\end{equation}
In particular the line $\psi = W_{-2,-1} = e^{-i \int \left(2 b +a  \right) }$ is transparent with charge $-1$ and fermionic statistics and so corresponds to the worldline of the original electron. Overall this TQFT has $2(2n+1)$ unique lines up to transparent lines of integer spin, which we can take to be generated by a fundamental anyonic line with $n_a=1$ and $n_b=0$. This spectrum exactly matches that of the known Jain sequence states. 

As we discuss in Appendix~\ref{A:oneForm} this TQFT has a $\mathbb{Z}_{|2n+1|}$ one-form symmetry on a torus which shifts $(a,b)$ by fractional large gauge transformations. This symmetry is the low-energy version of the magnetic translation symmetry of non-relativistic interacting fermions at filling fraction $\frac{n}{2n+1}$, and in the low-energy Chern-Simons theory it permutes the various anyonic Wilson lines while leaving invariant the line $W_{-2,-1}$ describing the original electron. 

\subsection{HLR and non-relativistic limits}
\label{S:HLR}

The HLR theory~\cite{HLR} of the half-filled Landau level is described by a non-relativistic spinless fermion $\psi$ coupled to a gauge field with Lagrangian
\begin{equation}
	i \psi^{\dagger}(\partial_t + i c_0 -i A_0)\psi  - \frac{1}{2m}|(\partial_i + i c_i - i A_i)\psi|^2 +\frac{1}{8\pi}cdc\,.
\end{equation}
Here $\psi$ is a non-relativistic composite fermion. Note the improperly quantized Chern-Simons term for $c$. This model is ill-defined with $c$ a standard $U(1)$ field. However, in this context there is a simple remedy pointed out in Appendix C of~\cite{Seiberg_2016}. Here $c_0$ acts as a Lagrange multiplier enforcing the flux attachment
\begin{equation}
	\rho_{\psi} = \frac{1}{4\pi} \epsilon^{ij}\partial_i c_j\,,
\end{equation}
so that the charge obeys $Q_{\psi} = \frac{N_c}{2}$ with $N_c$ the flux of $c$. Since the charge is integer we see that this constraint is consistent only when the flux of $c$ is even. With this in mind we can replace $c\to 2c$ so that the constraint reads $Q_{\psi} =N_c$ and take the new $c$ to be a standard $U(1)$ field. The redefined theory is
\begin{equation}
	i \psi^{\dagger}(\partial_t + 2i c_0 - i A_0) \psi - \frac{1}{2m}|(\partial_i + 2i c_i - i A_i)\psi|^2 + \frac{2}{4\pi}cdc\,,
\end{equation}
which is well-defined and moreover obeys the spin-charge relation. 

Now consider the gapped state where we fill $n$ Landau levels of $\psi$. Integrating it out we effectively replace the fermion action by a Chern-Simons term for $2c+A$, leading to the low-energy Lagrangian
\begin{equation}
	\frac{n}{4\pi}(2c-A)d(2c-A) + \frac{2}{4\pi}cdc\,.
\end{equation}
This is a $U(1)_{-2(2n+1)}$ Chern-Simons theory. The filling fraction corresponding to this state is $\frac{n}{2n+1}$, however the TQFT differs from the standard one. One way to see this is that it has a larger one-form symmetry on a torus, $\mathbb{Z}_{2(2n+1)}$. Another is to compute the spectrum of non-trivial line operators. A Wilson line $\exp\left( i n_c\int_C  c\right)$ has spin (self-statistics) and charge
\begin{equation}
	S(n_c) = - \frac{n_c^2}{2(2n+1)}\,, \qquad Q(n_c) = \frac{n n_c}{2n+1}\,.
\end{equation}
Like the standard construction there are $2(2n+1)$ equivalent lines up to integer spin and even charge, however a comparison of the spins and charges here with those of~\eqref{E:basicJainLines} (say with $n_b = 0$ there) shows that the spectra differ. Relatedly this model lacks a transparent electron: transparent lines are multiples of the basic one with $n_c = 2(2n+1)$ and so integer spin $S =-2(2n+1)$ and even charge $Q = 2n$. 

An improved version of the HLR theory has been known in the condensed matter literature and was also described in Appendix C of~\cite{Seiberg_2016}. Like the HLR theory it does not respect particle-hole symmetry and it reads
\begin{equation}
\label{E:HLRtake2}
	i \psi^{\dagger}(\partial_t - i a_0)\psi - \frac{1}{2m}|(\partial_i - i a_i)\psi|^2 - \frac{1}{2\pi} a db - \frac{2}{4\pi}bdb + \frac{1}{2\pi} bdA\,.
\end{equation}
Here $b$ is a $U(1)$ field and, as noted there, $a$ should be understood as a spin$_c$ field. With these definitions this theory respects the spin-charge relation. Like the HLR theory it possesses a metallic Fermi liquid state at half-filling, as well as gapped states where $n-1$ Landau levels of $\psi$ are completely filled. These are described by the TQFT (here we can take $g$ to be the flat metric)
\begin{equation}
	n CS(a,g) - \frac{1}{2\pi} a db - \frac{2}{4\pi}bdb + \frac{1}{2\pi} bdA\,.
\end{equation}
But this is nothing more than the description~\eqref{eq:CS_Jain} obtained from depleting $n$ Landau levels of our Dirac composite fermion. We have already shown that this TQFT is equivalent to the standard one on a space without boundary.

There may be an even more direct connection between the SSWW theory and the improved version~\eqref{E:HLRtake2} of the HLR theory. We argue here that this improved version is a non-relativistic limit of the SSWW theory, similar to how the standard HLR theory has been argued to be a non-relativistic limit of Son's theory of the half-filled level~\cite{Son_2015}. Our argument emulates Son's. We deform the SSWW theory by a large $\mathsf{CT}$-violating Dirac mass $m$ for the composite fermion and study a regime where the effective chemical potential $a_0$ for it is tuned so that $a_0>|m|$ but $|a_0-m|\ll |m|$. The effective Lagrangian is
\begin{equation}
\label{E:CFwithmass}
	i \bar{\chi}\slashed{D}_a \chi - m \bar{\chi}\chi - \frac{1}{2\pi}adb - \frac{2}{4\pi}bdb + \frac{1}{2\pi} bdA\,.
\end{equation}
In this regime one of the two components of $\chi$ is heavy while the other, call it $\psi$, remains light and it should be retained in the effective description. Integrating out the heavy component  $i \bar{\chi}\slashed{D}_a \chi$ gets replaced by an approximate non-relativistic action for $\psi$. With our convention for the Dirac operator this includes a radiatively induced Chern-Simons term $-CS(a,g)$ when $m>0$ but for $m<0$ no such term is generated. For $m>0$ the resulting theory is nothing more than the improved version~\eqref{E:HLRtake2} of the HLR theory.

To the extent that we can trust the description~\eqref{E:CFwithmass} for any Dirac mass $m$ it has a metallic state at half filling described by a Fermi liquid coupled to Chern-Simons gauge fields. Because the charge of $\chi$ is fixed to be $\frac{N_{\phi}}{2}$ at this filling the Fermi momentum is not a function of the mass at tree level, however the Fermi velocity is. It would be interesting to calculate the Landau parameters for this model. On symmetry grounds we expect them to be a non-trivial function of the mass, with backscattering forbidden only at the particle-hole symmetric value $m=0$, in which case this model really describes a line of fixed points parameterized by $m$. There is a special point on this line, the SSWW theory at $m=0$, which respects particle-hole symmetry.  The Jain states arise from moving off this line, and the mass does not alter the low-energy TQFT description of those states. However the states with negative $n$ are readily understood from the Dirac composite fermion. Perhaps this is not surprising, since the improved version of the HLR theory can be thought of as coming from a limit where we zoom on a large positive mass $m$, after which the states of $\chi$ with negative $n$ are no longer accessible to us.

\section{Reapplying the duality and the full hierarchy}
\label{sec:main}

While the basic duality already elegantly explains a metallic state at half filling and the Jain states emanating from it, we can push this description even further. There are states of the Dirac composite fermion theory~\eqref{eq:duality} where the composite fermion feels a large effective magnetic field, forms Landau levels, and some number $n-1$ of them are completely filled with one partially filled. Assuming weak interlevel mixing, an assumption which we expect to hold when the effective field is strong compared to the scale of interactions, the effective description of this partially filled level is strongly coupled, since the only low energy scale is set by interactions. In fact it is quite similar to that of interacting electrons in the lowest Landau level. However now we claim that the duality \eqref{eq:duality} allows us to dualize the partially filled level into a new composite fermion theory. As we will see this new description has a metallic state at some other even denominator filling besides 1/2, depending on $n$, along with gapped Jain states emanating from it. This is analogous to the flux attachment procedure for non-relativistic composite fermions. 

Our starting point is to take both sides of the fundamental fermion-fermion duality~\eqref{eq:duality}, deform by a magnetic field for $A$, and tune the chemical potential so as to complete fill all Landau levels of the free fermion $\psi$ up through level $n-1$, and we then partially fill $n$. We claim that this leads to an infrared duality
\begin{equation}\label{eq:dualityn}
     i\bar{\psi}\slashed{D}_A\psi   + CS(A,g)\quad \xLongleftrightarrow{n~<~\nu~<~n+1} \quad i\bar{\chi}\slashed{D}_a\chi - \frac{1}{2\pi} a db - \frac{2}{4\pi} b  db + \frac{1}{2\pi} b  dA + n CS(A,g)\,.
\end{equation}
The additional level $n$ Chern-Simons term for $A$ on the right-hand side arises in order to match the local action that arises from integrating out the completely filled Landau levels of $\psi$. The ensuing effective description in the partially filled Landau level is similar to that of the zeroth Landau level as all levels are essentially the same. One can perform all of the same checks of \cite{Seiberg_2016} on this claimed duality. For example adding a fermion mass to the left-hand-side moves the spectrum to a gap on both sides of the duality, resulting in a trivial gapped state with effective Lagrangian $nCS(A,g)$ for positive mass, and $(n+1)CS(A,g)$ for negative mass. Deforming the right-hand-side by a large fermion mass for $\chi$ (large compared to any irrelevant operators but small compared to $\sqrt{B}$) we find a trivial phase described by $nCS(A,g)$ for negative mass and $(n+1)CS(A,g)$ for positive mass. We have in \eqref{eq:dualityn} ignored the shift in the Landau levels, or equivalently the coupling to the spatial spin connection. We refine the proposal~\eqref{eq:dualityn} to include these terms when we discuss the spin and shift in Subsection \ref{sec:spinshift}.

Now, entirely filling $n-1$ levels of the composite fermion on the right-hand-side of~\eqref{eq:duality}, partially filling the $n$th level, and dualizing the composite fermion we land on 
\begin{equation} \label{eq:level2}
\begin{aligned}
    ~i \bar{\eta} \slashed{D}_{a_2} \eta &- \frac{1}{2\pi}  a_2 d b_2 - \frac{2}{4\pi}  b_2  d b_2 + \frac{1}{2\pi }  b_2 da_1 + (n-1)CS(a_1,g) \\ & - \frac{1}{2\pi} a_1 db_1 - \frac{2}{4\pi}  b_1 d b_1 + \frac{1}{2\pi} b_1 d A \,.
\end{aligned}
\end{equation}
When $n=0$ we can integrate out all of the dynamical statistical fields in a consistent manner, starting with $a_1$, and get back to the left-hand-side of~\eqref{eq:dualityT}. However for general $n$ this model has a metallic state at even-denominator filling $\nu = \frac{2n-1}{4n}$, where the net flux of $a_2$ vanishes and the dualized fermion $\eta$ forms a large Fermi surface, and Jain states emanating from it. As in our discussion of the Dirac composite fermion in the last Section there are also implicitly further interactions, call them $\widetilde{O}$, which are generally irrelevant at the critical point alluded to above. As we will see all of these states occur in the range $\frac{n-1}{2n-1}\leq \nu \leq \frac{n}{2n+1}$ for $n>0$ and in the range $\frac{n-1}{2n-1}\geq \nu \geq \frac{n}{2n+1}$ for $n<0$.

The metallic state at filling $\nu = \frac{2n-1}{4n}$ is essentially a Fermi liquid coupled to Chern-Simons gauge fields $(a_1,b_2,a_2,b_2)$, all of which are effectively massive. Note that particle-hole symmetry acts nicely on these states, with $\nu \to 1-\nu$ corresponding to taking $n\to -n$.

As before, when the fermion $\eta$ feels a large flux we expect it to form Landau levels and for there to be gapped states where we some number of those levels are completely filled. We can identify the TQFT description of these states by replacing the fermions with a level $-m$ Chern-Simons term for $a_2$, where $m$ is the number of filled Landau levels. The resulting TQFT is described by a $K$ matrix and $t$ vector
\begin{equation} \label{eq:K2}
    K = \begin{pmatrix}
        2~ & 1 & 0 & 0 \\
        1 & ~\mathbf{1 - n}~ & ~-1~ & 0 \\
        0 & -1 & 2 & 1 \\
        0 & 0 & 1 & \mathbf{~-m}
    \end{pmatrix}\,, \qquad t = \begin{pmatrix} 1 \\ 0 \\ 0 \\ 0 \end{pmatrix}\,.
\end{equation}
These states have a filling $\nu = \frac{1}{2+\frac{1}{n-1+\frac{1}{2+\frac{1}{m}}}} = \frac{2mn - m + n - 1 }{4mn + 2n - 1}$, and for fixed $n$ describe a series of gapped states near $\nu = \frac{2n-1}{4n}$ as to be expected. As with our discussion of the Jain sequences with $\nu = \frac{n}{2n+1}$, this construction will break down at large $|m|$ where the interlevel spacing becomes comparable to the interaction scale.

This procedure of completely filling levels of the composite fermion, partially filling, and dualizing, can be iterated again and again, resulting in a ``hierarchy'' of composite fermion descriptions. The gapped states that arise after applying the duality $k$ times are described by the $K$ matrix and $t$ vector
\begin{equation} \label{eq:Kfull}
    K = \begin{pmatrix}
        2~ & 1 & 0 & 0 & 0 & \dotsm & 0   \\
        1 & ~\mathbf{1 - n_1}~ & ~-1~ & 0 & 0 & \dotsm& 0  \\
        0 & -1 & 2 & 1 & 0 & \dotsm & 0  \\
        0 & 0 & 1 & \mathbf{~1-n_2} & ~-1~ & \dotsm & 0 \\
        0 & 0 & 0 & ~-1~ & 2 & \ddots & \vdots \\
        \vdots & \vdots & \vdots & \vdots & \ddots & \ddots & 1 \\
        0 & 0 & 0 & 0 & \dotsm & 1 & ~\mathbf{-n_k}~
    \end{pmatrix}, \qquad \qquad \qquad
    t = \begin{pmatrix}
        1 \\ 0 \\ 0 \\ \vdots \\ 0
    \end{pmatrix}.
\end{equation}
This theory involves $k$ dynamical $U(1)$ fields and $k$ dynamical spin$_c$ fields, and so $K$ is a $2k \times 2k$ matrix. 

\begin{figure}
    \centering
    \includegraphics[width=0.7\linewidth]{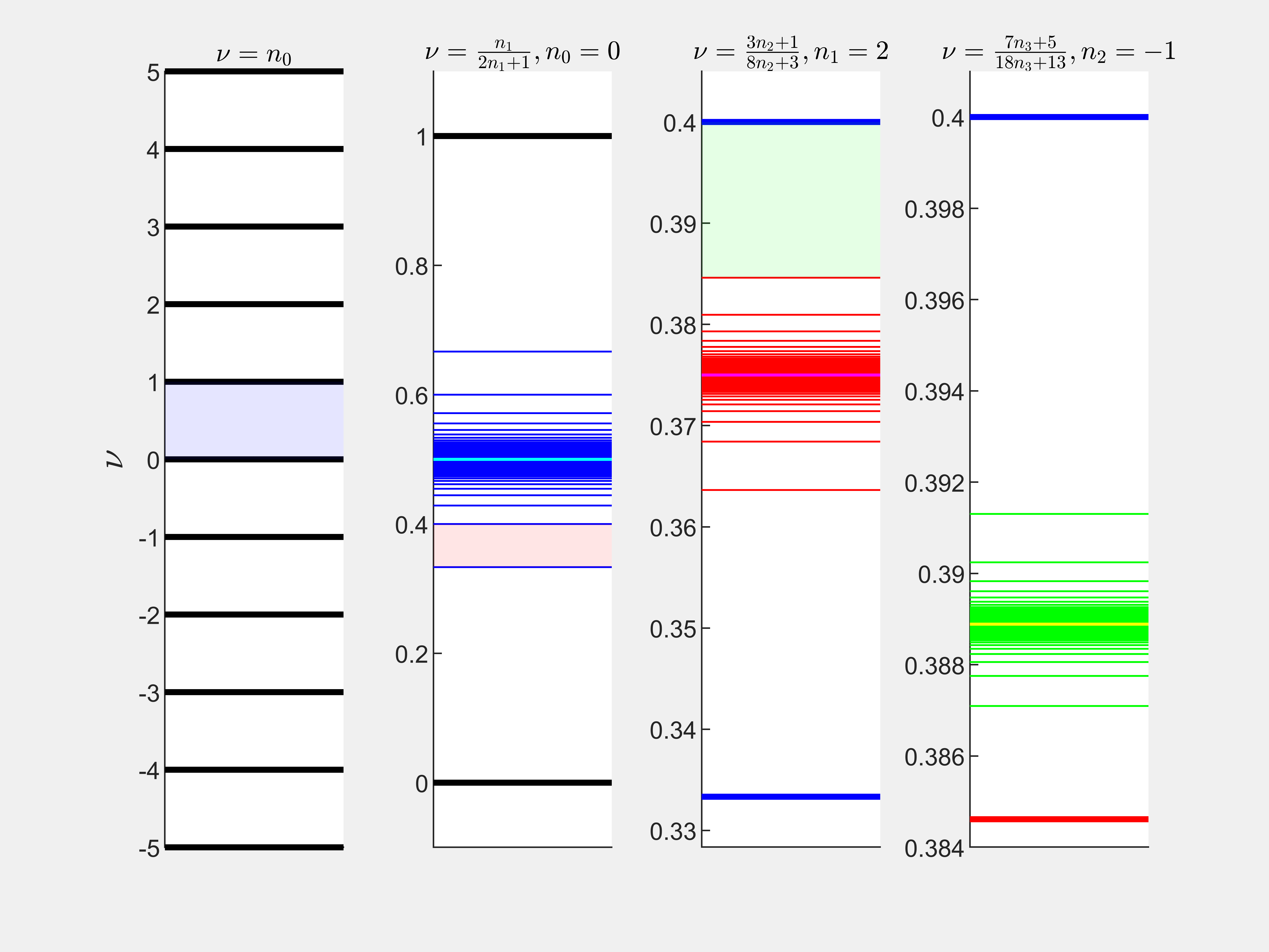}
    \caption{A depiction of the first few levels of the hierarchy. The black, blue, red and green lines represent gapped states in the zeroth, first, second and third levels respectively. The first is the integer quantum Hall effect. Subsequent levels focus on the shaded region in the previous level.}
    \label{fig:hierarchy}
\end{figure}

Calling $\nu(n_1,\ldots, n_k)$ the fillings of these gapped states, they satisfy the properties:
\begin{equation}
\begin{aligned}
    \nu(n_1,\ldots, n_{k-1},0) &= \nu(n_1,\ldots, n_{k-1}-1), \\ 
    \nu(n_1,\ldots, n_{k-1},-1) &= \nu(n_1,\ldots, n_{k-1}),\\
    \nu(n_1,\ldots, n_{k-1}-1) \leq \nu(n_1,&\ldots, n_{k_1},n_k) \leq \nu(n_1,\ldots, n_{k-1}),
\end{aligned} 
\end{equation}
with the equality in the last line being possible if and only if $n_k = 0$ or $n_k = -1$. The first two properties are simply algebraic statements on integrating out fields, while the last property is a consequence of the filling constraint in \eqref{eq:dualityn}. At each level of the hierarchy, or equivalently with each successive application of the duality, the gapped states lie in between two gapped states at one level higher. See Fig.~\ref{fig:hierarchy} for an illustration of this for the first few levels.

These properties immediately imply that the gapped state predicted by this duality with a given odd-denominator filling is unique, i.e. there are no other inequivalent gapped states with the same filling fraction predicted by the duality. 

\subsection{Existence of gapped states with arbitrary odd denominator fillings}

Here we will establish that for every rational filling with an odd denominator there is a unique gapped state predicted in the composite fermion duality construction. Uniqueness was already established before, so it remains to show the existence of a gapped state for each rational filling. We do this by induction. By particle-hole symmetry we can specialize to states with filling $\nu <\frac{1}{2}$. For filling $p/q$ this means that $p\leq \frac{q-1}{2}$ or equivalently $q \geq 2p+1$. This bound is saturated by the Jain states $\nu = \frac{p}{2p+1}$ which are produced by the first level of our hierarchy. These states will form the base case of the induction argument. 

Now let us assume that for fixed numerator $p$ that we have a construction for states with filling $p/q$ for all possible $q \leq m$ for some odd integer $m\geq 2p+1$. We would like to show that a state with filling $\frac{p}{m+2}$ also exists. To do so we use that the filling fraction corresponding to the $K$ matrix \eqref{eq:Kfull} can be represented as a continued fraction,
\begin{equation} \label{eq:nufrac}
    \nu(n_1,\ldots,n_k) = \frac{1}{2 + \frac{1}{n_1 -1 + \frac{1}{2 + \frac{1}{n_2 - 1 + \frac{1}{\ldots + \frac{1}{n_k}}} }}} = \frac{1}{2 + \frac{1}{n_1-1 + \nu(n_2,\ldots,n_k)}}\,.
\end{equation}
The filling $p/m$ can then be represented as $\nu(n_2,\ldots,n_k)$ for some integers $\{n_2,\ldots,n_k\}$. At the next level of the hierarchy there is a state with $n_1=1$ which has filling $\nu = \frac{1}{2+\frac{1}{0+\frac{1}{\nu(n_2,\ldots,n_k)}}} = \frac{p}{m+2}$ as desired. Repeating this argument demonstrates that the existence of states with filling $\frac{p}{q}$ of fixed numerator $p$ and all possible $q\geq 2m+1$.

To complete the proof we need only demonstrate that the initial assumption is true for the smallest possible value of $m$, $2p+1$ and all $p$. For fixed $p$ there is a unique such state with $\nu <1/2$, the Jain state $\frac{p}{2p+1}$, which we have already found for all $p$.

\subsection{Shift and spin} \label{sec:spinshift}

Up to now we have ignored the coupling to the spatial spin connection due to the relativistic shift in the degeneracy of the Landau levels. However, to correctly obtain the shift given in \eqref{eq:intQH} we have to amend the duality \eqref{eq:dualityn} to include Wen-Zee terms that account for this shift. To reproduce the correct effective action when we gap out both sides of the duality with fermion masses we infer that on a non-trivial gravitational background the duality reads
\begin{align}
\label{eq:dualityn+shift}
     i\bar{\psi}\slashed{D}_A\psi +CS(A,g) \quad \xLongleftrightarrow{n~<~\nu~<~n+1} \quad &i\bar{\chi}\slashed{D}_a\chi - \frac{1}{2\pi} adb  - \frac{2}{4\pi} b  db + \frac{1}{2\pi} b  dA + \frac{|n|}{2\pi}  b d \omega' 
     \\
     \nonumber
     &+ nCS(A,g) + \frac{|n|(n-1)}{4\pi} A  d\omega'  + \frac{n(n-1)(2n-1)}{24\pi} \omega' d\omega'\,.
\end{align}
The second line is a local effective action arising from integrating out the completely filled Landau levels of the fermion $\psi$.\footnote{Recall that we are measuring the shift relative to the gap between the zeroth and $-1$ Landau levels, so for a partially filled $n)$th level of $\psi$ we completely fill all levels through $n-1$.}  The coupling of $b$ to the spatial spin connection is then fixed by demanding that both sides produce the same effective action when gapping out both sides with a fermion mass, with a positive mass for $\psi$ corresponding to a negative mass for $\chi$ and vice versa.

We can calculate the shift of the shift for these states using the standard formula \cite{Wen_Zee_92,Wen_1995}:
\begin{equation} \label{eq:Shift}
    \mathcal{S} = \frac{2 t^T K^{-1} s}{\nu} = \frac{2 t^T K^{-1} s}{t^T K^{-1} t}\,,
\end{equation}
where the spin vector $s$ would be in our case
\begin{equation} \label{eq:Spin}
    s = \begin{pmatrix}
    1/2~ & \frac{n_1|n_1-1|}{2}~ & |n_1|~ & \frac{n_2|n_2-1|}{2} ~& |n_2|~ & \dots~ & |n_{k-1}|~ & \frac{(n_k+1)|n_k|}{2}
    \end{pmatrix}^T \,.
\end{equation}
The $1/2$ in the first term comes from the background spin$_c$ field $A$, as was the case in the shift for the Jain sequence. Equivalently one can think of this $1/2$ as a correction to the relativistic shift to get the non-relativistic shift, similar to \cite{Son_2015}. 

Using equations \eqref{eq:Spin} and \eqref{eq:Shift}, the shift on the first two levels of the hierarchy is
\begin{equation}
\begin{aligned}
    \mathcal{S}_1 &= |n_1| + 1 + \text{sgn}(n_1)\,,\\
    \mathcal{S}_2 &= 1 + \frac{n_1|n_1-1|+2n_2 n_1 (|n_1-1|+1) + |n_2| + n_2| n_2|}{2n_1n_2 + n_1 - n_2 -1}\,.
\end{aligned}
\end{equation}
The first line is the well-known shift of the Jain sequence derived previously.

\subsection{A simplification for states around $\nu =\frac{1}{2n}$}

Consider the second level of this hierarchy at the first Landau level, that is Eq.~\eqref{eq:level2} with $n=1$. For this special case $a_1$ appears linearly and so we can integrate it out, setting $b_1 = b_2 + (\text{gauge})$. This gives us after a relabeling $b_2\to b$ and $a_2 \to a$,
\begin{equation} 
\label{eq:1/4}
    i \bar{\eta} \slashed{D}_{a} \eta - \frac{1}{2\pi}  a  d b - \frac{4}{4\pi} b  d b + \frac{1}{2\pi} b  d A + \frac{1}{2\pi}b d\omega' \,.
\end{equation}
This model possesses a gapless state at $\frac{1}{4}$ filling. Dropping the last term, it has appeared previously in the literature~ \cite{Goldman_2018,Wang_2019,Nguyen_2021} as a proposal for a description of that state. Iterating this procedure $n$ times -- i.e. fully filling the lowest Landau level, partially filling the first, dualizing using~\eqref{eq:dualityn+shift}, and then integrating out the Lagrange multiplier $a_1$ -- we find the same model proposed by Goldman and Fradkin~\cite{Goldman_2018} for gapless states at $\nu = \frac{1}{2n}$,
\begin{equation} \label{eq:1/2n}
    i \bar{\eta} \slashed{D}_{a} \eta - \frac{1}{2\pi} a  d b - \frac{2n}{4\pi}b d b + \frac{1}{2\pi}  b d A + \frac{n-1}{2\pi} b d \omega'\,.
\end{equation}
The last term, the coupling of $b$ to the spatial spin connection was absent in that work and is new to this one, a consequence of the duality~\eqref{eq:dualityn+shift}. The TQFT description of the gapped states emanating from this sequence comes from simply replacing the composite fermions with a Chern-Simons term at level $m$, where all Landau levels less than $m-1$ are filled. These gapped states have a filling $\nu = \frac{m}{2mn+1}$, a $K$ matrix $K = \begin{pmatrix}
    2n & 1 \\ 1 & \mathbf{-m}
\end{pmatrix}$, and a $t$ vector $t = \begin{pmatrix} 1 & 0 \end{pmatrix}^{\rm T}$. Note that here $n$ denotes the number of times we use dualize and so $n>0$ while $m$ can be arbitrary.

The spin vector and shift of these theories can be found using the computations in the previous subsection to be
\begin{equation}
    s = \begin{pmatrix}
        n - \frac{1}{2}~~ & \frac{|m|(m+1)}{2}
    \end{pmatrix}^T, \qquad \qquad \mathcal{S} = |m| + 2 n - 1 + \text{sgn}(m)\,.
\end{equation}
This shift matches the well known result of the Jain sequence near $\nu = \frac{1}{2n}$. We want to contrast this result with previous computations in~\cite{Goldman_2018,Nguyen_2021} of the shift for these theories. Those authors noticed that the correct shift does not arise solely from the contribution of partially filling the composite fermions. While the authors of \cite{Goldman_2018} corrected the shift by redefining the background field ($A \to A + (n-1)\omega'$), the authors of \cite{Nguyen_2021} did so by introducing an extra spin-2 massive degree of freedom, a ``Haldane mode,'' which below the scale of its mass reduces to the last term in~\eqref{eq:1/2n}. Here we showed that this term indeed follows from the duality on a weakly curved background. Our analysis is valid in the deep infrared, and therefore is unable to determine if this last term $\frac{1}{2\pi} bd\omega'$ arises from a ``Haldane mode'' as in~\cite{Nguyen_2021}.

While the description of the $\nu = \frac{1}{2n}$ states is quite simple, in terms of a fermion coupled to only $a$ and $b$, the particle hole conjugate at $\nu = \frac{2n-1}{2n}$ is more complicated. For example, the $3/4$ state comes from taking~\eqref{eq:level2} with $n=-1$, describing a fermion coupled to two spin$_c$ fields $a_1$ and $a_2$ and two $U(1)$ fields $b_1$ and $b_2$. More generally the state at $\nu = \frac{2n-1}{2n}$ is described by a fermion coupled to $n$ spin$_c$ fields and $n$ $U(1)$ fields.

\subsection{Equivalence to the standard hierarchy construction}

Just as we did for the Jain sequence in Subsection \ref{sec:Jain_K}, we can map the TQFT describing the more general states given by \eqref{eq:Kfull} to the expected TQFT description coming from the hierarchical version~\cite{PhysRevLett.65.1502} of Jain's construction~\cite{PhysRevLett.63.199}. This in turn is equivalent to a TQFT description coming from the Haldane-Halperin hierarchy~\cite{PhysRevLett.51.605,PhysRevLett.52.1583}. We do so on a spacetime without boundary and proceed in the same way as before, trading the Chern-Simons terms for the spin$_c$ fields by integrating in a number of $U(1)_{\pm 1}$ fields. For simplicity we will only show how to map the $K$ matrices and $t$ vectors, though one can also map the spin vectors to those of the hierarchy construction in a similar fashion.

At the second level of our hierarchy we have after integrating in $|n-1|$ $U(1)$ fields $c_i$ and $|m|$ $U(1)$ fields $\tilde{c}_j$,
\begin{equation}
    \begin{aligned}
    \label{E:JainTQFTsecond}
    -\frac{\text{sgn}(m)}{4\pi} \sum_{j=1}^{|m|} \tilde{c}_j  d\tilde{c}_j & - \frac{2}{4\pi}   b_2 d b_2- \frac{1}{2\pi} b_2 dA- \frac{1}{2\pi} a_2  d \left(b_2 - \sum_{j=1}^{|m|}\tilde{c}_j \right) 
    \\ 
   & -\frac{\text{sgn}(n-1)}{4\pi} \sum_{i=1}^{|n-1|} c_i  dc_i+ \frac{1}{2\pi}\sum_{i=1}^{|n-1|}c_i dA
   \\
   & \qquad - \frac{2}{4\pi}  b_1  d b_1 \,.
   \\
   & \qquad \qquad -\frac{1}{2\pi}a_1  d\left(b_1-b_2 -\sum_{i=1}^{|n-1|} c_i\right)\,.
\end{aligned}
\end{equation}
We have deliberately chosen this unusual formatting. Shortly, it will allow us to show its equivalence not merely with the standard TQFT description using composite fermions, but with the method used to obtain it.

To arrive at this expression we used that $a_1$ enforces the constraint $db_1 =db_2 + \sum_{i=1}^{|n-1|}dc_i$ to rewrite $\frac{1}{2\pi}b_1 dA$ in terms of other couplings to the background field $A$. Now integrating out the spin$_c$ fields $a_1$ and $a_2$ leads to the constraints $b_2 = \sum_{j=1}^{|m|} \tilde{c}_j + (\text{gauge})$ and $b_1 = \sum_{j=1}^{|m|} \tilde{c}_j + \sum_{i=1}^{|n-1|} c_i + (\text{gauge})$. The resulting $K$ matrix and $t$ vector for the $U(1)$ fields $\tilde{c}_j$ and $c_i$ are 
\begin{align}
\begin{split}
    \label{eq:K2inU1}
    	K^{(2)} &= \begin{pmatrix}
        \text{sgn}(m)I_{|m|} + 4 J_{|m|} & 2 J_{|m| \times |n-1|} \\
        2 J_{|n-1| \times |m|} & \text{sgn}(n-1) I_{|n-1|} + 2 J_{|n-1|}
    \end{pmatrix} 
    \\
    & = \begin{pmatrix}
        \text{sgn}(m)I_{|m|} + 2 J_{|m|} & 0 \\
        0 & \text{sgn}(n-1) I_{|n-1|}
    \end{pmatrix} + 2 J_{|m| + |n -1|}\,,
    \\
    t &= \begin{pmatrix}
        1_{|m|} \\ 1_{|n-1|},
    \end{pmatrix} = 1_{|m|+|n-1|}\,,
\end{split}
\end{align}
where, as before, $I_n$ is the $n \times n$ identity matrix, $J_n$ is the $n \times n$ matrix with all entries being $1$, and $1_n$ is a vector of ones with length $n$. This $K$ matrix can be thought of in the following way. We begin by adding three decoupled TQFTs together. The first is the one describing the Jain states given by $K^{(1)} = \text{sgn}(m)I_{|m|} + 2J_{|m|}$ together with $t^{(1)} = 1_{|m|}$ in terms of $U(1)$ fields $\tilde{c}_j$. The second describes the integer quantum Hall state at level $|n-1|$, i.e. it has a $K$ matrix $K^{(2)} = \text{sgn}(n-1)I_{|n-1|}$ and $t$ vector, $t^{(2)} = 1_{|n-1|}$ described in terms of $U(1)$ fields $c_i$. The third is a $U(1)_2$ Chern-Simons theory $-\frac{2}{4\pi} bdb$ with no coupling to the background field. Then one couples them together by introducing a Lagrange multiplier field setting the current constructed from $b$ to the sum of the charge currents from the first and second theories, i.e. it enforces the condition
\begin{equation}
	\frac{1}{2\pi}db =  \frac{1}{2\pi}\left( \sum_{j=1}^{|m|} t^{(1)}_j d\tilde{c}_j + \sum_{i=1}^{|n-1|} t^{(2)}_i dc_i\right)\,.
\end{equation}
In colloquial terms two units of flux have been attached to the total charge current. This is equivalent to the TQFT in~\eqref{E:JainTQFTsecond}. To see this note that the first line of that expression is, after integrating out the Lagrange multiplier $a_2$, the TQFT describing of the Jain state. The second line describes the integer Hall state, and the third is the $U(1)_2$ theory. In the last line $a_1$ acts as a Lagrange multiplier soldering the currents of the three TQFTs together. But this presentation is nothing more than the construction of the Jain hierarchy described in \cite{wen1992classification}.

In general, the $K$ matrix at the $k$th level in this hierarchy $\nu(n_1,\ldots ,n_k)$ can be constructed in a similar way as
\begin{equation}
\label{E:JainHierarchy}
    K^{(k)} = \begin{pmatrix}
        K^{(k-1)} & 0 \\
        0 & \text{sgn}(n_k-1) I_{|n_k-1|}
    \end{pmatrix} + 2 J\,, \qquad t = 1\,,
\end{equation}
where $K^{(k-1)}$ is the $K$ matrix of a state in the $k-1$th level described by a set of nonzero integers $\{n_1,\ldots,n_{k-1}\}$. We can interpret this TQFT in the following way. At each level in the hierarchy we take a TQFT from the previous level and add it to an integer Hall state, and then couple them together by adding the matrix of all twos. This last step is equivalent to adding a decoupled $U(1)_2$ theory $-\frac{2}{4\pi}bdb$ and then enforcing a constraint setting the current $\frac{1}{2\pi}db$ equal to the sum of the charge currents of the integer Hall state and the seed in the $k-1$th level. Or, it describes the attachment of two units of flux to the total charge current.

Again, these are the exact steps described in \cite{wen1992classification} for the Jain version of the hierarchy construction, except that \cite{wen1992classification} allows the addition of $2J$ to be an addition of $2 p J$ for any arbitrary positive integer $p$. This is also contained in our construction although at level $k-1+p$. This state is described by the set of integers $\{n_1,\ldots n_{k-1},\underbrace{1,1,\ldots,1}_{p-1},n_k\}$. The effect of the 1s is similar to that described in the last Subsection describing gapless states at $\nu = \frac{1}{2n}$. The end result is a TQFT that can be written as the sum of three terms -- the seed in the $k-1$th level, the integer Hall state with filling $|n_k-1|$, and a $U(1)_{2p}$ theory $-\frac{2p}{4\pi}bdb$ -- together with a Lagrange multiplier field $a$ equating the current $\frac{1}{2\pi}db$ with the sum of the charge currents of the other two sectors. This effectively adds $2pJ$ instead of $2J$ in~\eqref{E:JainHierarchy}, or attaches $2p$ units of flux. Note that the here $p>0$ since it corresponds to a number of times that we implement the duality~\eqref{eq:dualityn}.

\subsection{Non-relativistic limits and gaps}

As with our discussion in Subsection~\ref{S:HLR} the Dirac composite fermion theories discussed above can be deformed by Dirac masses. In that context particle-hole symmetry forbids such a mass term but here there does not seem to be a principled reason to not include it. For the Fermi surfaces at even-denominator filling obtained in this Section a fermion mass leaves the existence of the Fermi surface intact while altering the Landau parameters, while also leaving along the Chern-Simons description of the Jain states emanating from them. By taking large masses we can zoom in on non-relativistic versions of these Dirac composite fermions, leading to improved versions of HLR-like theories for the Jain hierarchy. For example, at the second level of the hierarchy after taking a large negative mass $m$ and tuning $(a_2)_0=|m| + \tilde{a}_2$ for a small positive fluctuation $\tilde{a}_2$ we arrive at an effective description
\begin{align}
\nonumber
	i \eta^{\dagger} (\partial_t - i (\tilde{a}_2)_0)\eta &- \frac{1}{2|m|} |(\partial_i - i (a_2)_i)\eta|^2 - \frac{1}{2\pi} a_2 db_2 - \frac{2}{4\pi}b_2db_2 + \frac{1}{2\pi}b_2 da_1 + (n-1)CS(a_1,g)
	\\
	& -\frac{1}{2\pi} a_1 db_1 - \frac{2}{4\pi}b_1db_1 + \frac{1}{2\pi}b_1 dA\,.
\end{align}

When the composite fermion feels a magnetic field we expect it to form Landau levels whose spacings depend on the strength of the effective field and on the fermion mass. In the non-relativistic limit these spacings are equal. Here we discuss the size of the gaps under the scenario that the mass is small.

\begin{figure}
    \centering
    \includegraphics[width=0.8\linewidth]{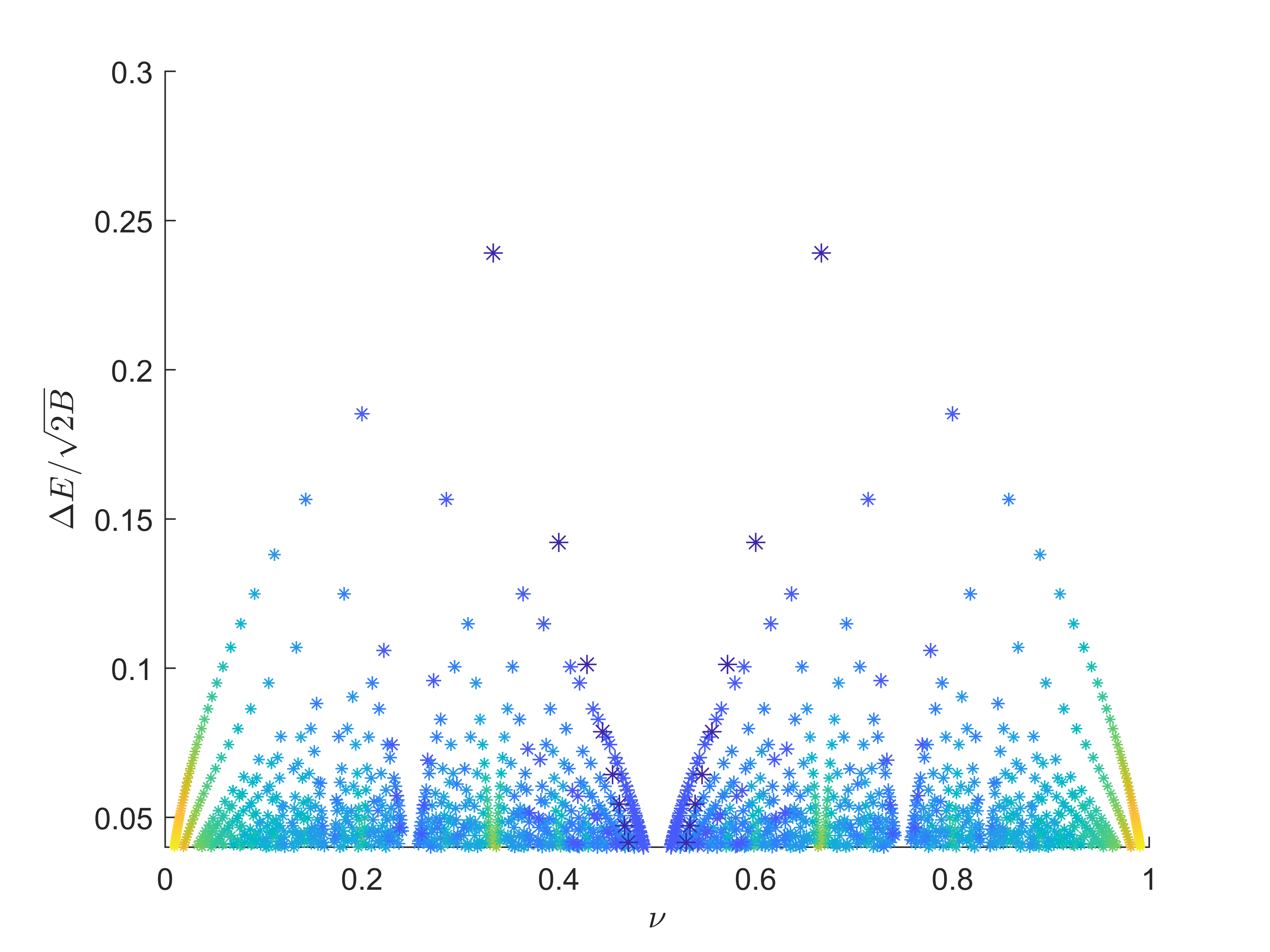}
    \caption{The single particle energy gap for the gapped states in the hierarchy in the scenario that the composite fermion has no Dirac mass. The marker size and color correspond to the level of the state in the hierarchy, i.e. the number of times the duality was used. The largest markers and dark blue color correspond to the first level or Jain sequence states. The smaller and yellower markers are further down in the hierarchy.}
    \label{fig:gap}
\end{figure}

In the gapped states at the $k$th level of the hierarchy the composite fermions completely fill all Landau levels up to $n_k-1$, so the lowest energy excited states would be ones where a single composite fermion is moved from the last filled Landau level to one higher. The energy of these Landau levels is $E_{n_k} = \text{sgn}(n_k)\sqrt{2 |B_{a_k} n_k|}$, where $B_{a_k} = B (K^{-1})_{1,2k}$ is the magnetic field felt by the composite fermions. One can check that $|B_{a_k}| = |B/\det(K)|$, or in other words if we consider a state with rational filling $\nu = p/q$ then the magnetic field felt by the composite fermions is $B/q$. This statement only depends on the Chern-Simons description of the gapped state and not on whether the microscopic fermion has a mass or not. The gap to excite a single composite fermion is then
\begin{equation}\label{eq:Egap}
    \Delta E = \sqrt{\frac{2B}{q}}\left|\sqrt{|n_k+1|} - \sqrt{|n_k|} \right|.
\end{equation}
We see that this gap both decreases with $q$ as well as with $n_k$. A figure of the gap size for various gapped states with different filling fractions is given in Fig.~\ref{fig:gap}. 

For the Jain sequence, the gap is
\begin{equation}\label{eq:EgapJain}
    \Delta E = \sqrt{\frac{2B}{|2n+1|}}\left|\sqrt{|n+1|} - \sqrt{|n|} \right| \approx \frac{\sqrt{B}}{|2n+1|} + O(n^{-3}).
\end{equation}
This is in agreement with the gap predicted by the non-relativistic composite fermion picture, as well as with experiments \cite{jain2007composite}. We note that at small $n$ the predictions here deviate more from the non-relativistic composite fermion predictions, and this deviation may be testable. 

The single particle gap is not the full picture of the validity of the gapped state description, as really the regime of validity for the gapped states is when there is no Landau level mixing. Once there is significant mixing between Landau levels the gapped approximation breaks down, and the composite fermion system will be in the Fermi liquid regime. This happens when the interaction scale is larger than the magnetic field felt by the composite fermions, or in other words the composite fermions will form a large Fermi surface with weak perturbations when $\Lambda_{\text{interaction}} \gg B/q$, while they will form and fill Landau levels with some weak perturbations when $\Lambda_{\text{interaction}} \ll B/q$. Thus the prediction of which gapped states are most dominant, and so most likely to be seen, coincides with observations \cite{Bergholtz_QHTT,Hansson_QHrev}.

\section{States at even denominator fillings} 
\label{sec:even}

At even denominator fillings the duality predicts gapless metallic states, assuming there are no superconducting instabilities, similar to the one predicted at $\nu = 1/2$. At these fillings the magnetic field felt by the composite fermions is zero, while the average density of composite fermions at the $k$th level of the hierarchy is $2 \pi \rho_\chi = db_k $. We can compute $db_k$ and the filling fraction by considering the $K$ matrix without the last entries, as $da_k = 0$,
\begin{equation} \label{eq:Ktildefull}
    \tilde{K} = \begin{pmatrix}
        2~ & 1 & 0 & 0 & 0 & \dotsm & 0 & 0   \\
        1 & ~\mathbf{1 - n_1}~ & ~-1~ & 0 & 0 & \dotsm& 0 & 0  \\
        0 & -1 & 2 & 1 & 0 & \dotsm & 0 & 0  \\
        0 & 0 & 1 & \mathbf{~1-n_2} & ~-1~ & \dotsm & 0 & 0  \\
        0 & 0 & 0 & ~-1~ & 2 & \ddots & \vdots & \vdots \\
        \vdots & \vdots & \vdots & \vdots & \ddots & \ddots & 1 & 0 \\
        0 & 0 & 0 & 0 & \dotsm & 1 & ~\mathbf{1-n_{k-1}}~ & -1 \\
        0 & 0 & 0 & 0 & \dotsm & 0 & -1 & 2
    \end{pmatrix} \,.
\end{equation}
Then we can compute the filling fraction and $db_k$ as
\begin{equation}
    \nu = \left(\tilde{K}^{-1}\right)_{1,1}, \qquad \qquad
    db_k = \left(\tilde{K}^{-1}\right)_{1,2k-1}B .
\end{equation}
For a filling $\nu =p/q $ with $q = \det(\tilde{K})$ we have $|db_k| = B/q$. Thus the duality predicts that the density of composite electrons in the metallic states are
\begin{equation} \label{eq:rhochi}
    2\pi \rho_\chi = \frac{B}{q}.
\end{equation}
This is a nontrivial testable prediction that the sizes of the Fermi surface at many different fillings should coincide. In particular the Fermi momentum at $\nu = \frac{1}{8},\frac{3}{8},\frac{5}{8},$ and $\frac{7}{8}$ should all be identical, and one quarter the size of the Fermi momentum at $\nu = \frac{1}{2}$.

If a superconducting instability does occur resulting in BCS pairing of the composite electrons, the resulting gapped state can also be predicted. As in \cite{seiberg2016gappedboundaryphasestopological,Seiberg_2016}, the gapped state of the $s$-wave BCS condensation is described by gauging $a_k$ to be a $\ZZ_2$ field, then tensoring with an Ising sector, and finally modding out by a diagonal $\ZZ_2$. The resulting theory is typically broken up into an Abelian sector, and the Ising sector, which are coupled by modding out a diagonal $\ZZ_2$ 1-form symmetry. 

For the state near $\nu = 1/2$ the Abelian sector is a $U(1)_{8}$ theory:
\begin{equation}
\begin{aligned}
    L_{\rm eff} & = \frac{2}{4\pi} c da - \frac{1}{2\pi}  a  db - \frac{2}{4\pi} b db + \frac{1}{2\pi}  b dA  \\
    &\cong - \frac{8}{4\pi} c  dc + \frac{2}{2\pi}  c  dA \, ,
\end{aligned}
\end{equation}
and the full state resulting from adding the Ising sector and modding out by the $\ZZ_2$ subgroup is the T-Pfaffian or PH-Pfaffian state given by $\left(\text{Ising}\times U(1)_{8}\right)/\ZZ_2$ \cite{Lian_2018,Hsin_2020}.

We can repeat the same procedure for other even denominator states in the Hierarchy. For example, the gapped states with filling $\nu = \frac{1}{2n}$ will have an Abelian sector of $U(1)_{8n}$, and the resulting state is $\left(\text{Ising}\times U(1)_{8n}\right)/\ZZ_2$. This is a candidate state for fillings $\nu = 2 + \frac{1}{2n}$. We can also repeat the procedure for more general hierarchy states, which will result in a more general Abelian sector with a $K$ matrix of the form
\begin{equation}
    K = \begin{pmatrix}
        2~ & 1 & 0 & 0 & 0 & \dotsm & 0 & 0   \\
        1 & ~\mathbf{1 - n_1}~ & ~-1~ & 0 & 0 & \dotsm& 0 & 0  \\
        0 & -1 & 2 & 1 & 0 & \dotsm & 0 & 0  \\
        0 & 0 & 1 & \mathbf{~1-n_2} & ~-1~ & \dotsm & 0 & 0  \\
        0 & 0 & 0 & ~-1~ & 2 & \ddots & \vdots & \vdots \\
        \vdots & \vdots & \vdots & \vdots & \ddots & \ddots & 1 & 0 \\
        0 & 0 & 0 & 0 & \dotsm & 1 & ~\mathbf{1-n_{k-1}}~ & -2 \\
        0 & 0 & 0 & 0 & \dotsm & 0 & -2 & 8
    \end{pmatrix} .
\end{equation}
In particular we can take $k =2$ and $n_1 = 2$ to find a gapped state with filling $\nu = \frac{3}{8}$, which is a candidate non-Abelian state for the plateau observed at $\nu = \frac{19}{8}$ \cite{Xia_2004,Pan_2008}. 

\section{Summary and discussion}
\label{S:discussion}

We have used three-dimensional dualities to argue that a variant of Son's Dirac composite fermion~\cite{Son_2015} proposed by Seiberg, Senthil, Wang, and Witten (SSWW)~\cite{Seiberg_2016} gives a unified description of FQH states within the lowest Landau level. This single theory has a vast landscape of vacua and critical points, giving candidate descriptions for any rational filling fraction in the range $0<\nu <1$. Moreover, using duality we argued that the SSWW theory is a good description of the microscopic system of interest in fractional Hall physics, interacting electrons in the lowest Landau level. This unified picture has many interesting consequences and predictions which we derived throughout the text. Some highlights include:
\begin{enumerate}
    \item The TQFT description of the gapped states naturally involve dynamical spin$_c$ fields, alongside the dynamical $U(1)$ gauge fields. While these spin$_c$ fields can be traded for a collection of dynamical $U(1)$ fields, the description involving dynamical spin$_c$ fields is both simpler and more compact. For example, comparing the $K$ matrices at the second level of the hierarchy given in Eqs.~\eqref{eq:K2} and \eqref{eq:K2inU1}, the one involving dynamic spin$_c$ fields is much simpler. 

    \item There exists a unique gapped state for each rational filling fraction $\nu = p/q$ with $q$ odd within this construction. We showed how to derive it, including the $K$ matrix and $t$ vector. We also showed how to calculate the spin vector which describes the coupling to the spatial metric, and the shift of each state, both of which reproduce their expected values.

    \item The single particle excitation gap of these TQFT states is given by the energy required to excite a single composite Dirac fermion to the next Landau level, which is calculable and, when there is no mass for the composite fermion, given by Eq.~\eqref{eq:Egap}. Unlike the half-filled state and the Jain sequences emanating from it, for which particle-hole symmetry forbids such a mass, we do not have a principled reason to demand the Dirac composite fermion in the higher levels of the hierarchy in this paper to be massless, and should such masses exist they modify the form of the energy gaps.

    \item At even denominator filling there is a richer set of possibilities. One is a gapless Fermi liquid coupled to various Chern-Simons gauge fields, with the size of the Fermi surface determined from Eq.~\eqref{eq:rhochi}. That result implies that the size of the Fermi surface goes as $B/q$, so that for example metallic states at $1/8$ and $3/8$ filling have Fermi surfaces of equal area. This is another non-trivial prediction of the theory. There are also different possible gapped states at even denominator filling that arise from BCS pair instabilities of these Chern-Simons-Fermi liquids. Many gapped states are possible depending on the angular momentum of the pairing instability, all of which have non-abelian order. $s$-wave pairing for example leads to a non-abelian state with Ising TQFT and $U(1)$ Chern-Simons factors stitched together by a discrete quotient.

\end{enumerate}

In light of these predictions, we would like to call attention to some questions that our construction is well equipped to tackle, as well as some future directions worthy of exploration:
\begin{enumerate}
    \item The super-universality of the transitions between Hall plateaux seems evident from the description of gapped states as the IQHE of composite Dirac fermions. Nevertheless, this observation should be made more precise, especially when considering interactions. The effective field theory descriptions of our construction seem like an ideal settings to investigate these transitions.

    \item The role of disorder, and how it alters the composite Dirac fermion picture, is unclear. As disorder plays a large role in the IQHE, understanding how to incorporate disorder into this picture is a worthy pursuit. 

    \item Better control of how the energy gap changes due to interactions, sample width, and disorder, would be useful to more accurately compare the predictions from composite Dirac fermions to those of the standard non-relativistic composite fermion picture and experiments.

    \item While we can explain how gapped states at even denominator filling emerge from the super-conducting instability of the Fermi liquid description, it would be interesting to understand if and when these states can appear as non-perturbative vacua through the web of dualities. \cite{Goldman_2020} made some initial steps in this direction, but a unified understanding is still needed.
    
  \item Finally, while we have focused on the leading physics of the critical points of the SSWW theory it is straightforward to treat these Chern-Simons-Fermi liquids in effective field theory, including irrelevant corrections, to calculate response functions like the conductivity in a gradient expansion. Matching these corrections at the critical point will also lead to predictions for the gradient corrections to response in the Jain states emanating from these gapless states.
    
\end{enumerate}

\acknowledgments
It is a pleasure to thank Ofer Aharony, Luca Delacretaz, Bertrand Halperin, Andreas Karch, Allan MacDonald, Dam Son, Nathan Seiberg, Ramanjit Sohal, and Ashvin Vishwanath for valuable conversations. KJ was supported in part by an NSERC Discovery Grant. AR is supported by the U.S. Department of Energy under Grant No. DE-SC0022021 and a grant from the Simons Foundation (Grant 651678, AK).

\appendix

\section{Anomalous one-form symmetry of the fractional quantum Hall effect}
\label{A:oneForm}
It is illuminating to understand the 1-form symmetries of the TQFTs we found describing odd-denominator states, and their `t Hooft anomalies. As a review let us begin with analyzing the standard Abelian Chern-Simons theories with the Lagrangian 
\begin{equation}
    \mathcal{L} =  -\frac{K_{ij}}{4\pi}  b^i  d b^j + \frac{t_i}{2\pi} b^idA, 
\end{equation}
where $t\in \ZZ^n$ is a vector of integers that describes the coupling of the gauge fields to the electromagnetic current, and the symmetric matrix $K\in GL(n;\ZZ)$  describes the couplings between the various gauge fields. 

The one-form symmetry group of this theory consists of flat connections that leave the action invariant, but are not large gauge transformations \cite{Gaiotto_2015}. For simplicity let us put this theory on a spatial rectangular torus whose sides are of length $L_x$ and $L_y$. We also work in periodic imaginary time $\tau$. Then the large gauge transformation can be described by the shift $b^j \ra b^j + \xi^j \omega_{x,y}$, where $\omega_{x,y}$ is a properly normalized large gauge transformation winding around the $x$ or $y$ cycle and $\xi\in \ZZ^n$. For general $\xi\notin \ZZ^n$ the action under this transformation changes by $\Delta S = 2\pi i  n^i K_{ij}\xi^j$, where $n^j = \frac{1}{2\pi}\int_{C_2} db^j \in \ZZ$ are some integer-valued fluxes, with the 2-cycle $C_2$ looping around the $\tau-x$ or $\tau-y$ cycles. So $e^{-S_E}$ remains invariant if and only if $K\xi  \in \ZZ^n$. Overall this results in the symmetry group 
\begin{equation}
    G_{\text{one-form}} = \faktor{K^{-1}\ZZ^n}{\ZZ^n} \cong \faktor{\ZZ^n}{K\ZZ^n}\,.
\end{equation}
This group has size $|G_{\text{one-form}}| = |\det(K)|$, and the precise finite Abelian group can be found from the Smith normal form of $K$. In particular one can decompose $K = TBS$, with $T,S\in SL(n;\ZZ)$, and $B$ diagonal, with the diagonal entries in $B$ corresponding to the finite Abelian factors of $G_{\text{one-form}}$:
\begin{equation} \label{eq:g1form}
    G_{\text{one-form}} =\ZZ_{k_1} \times \ZZ_{k_2} \times \ldots \times \ZZ_{k_n} \,.
\end{equation}

We can also understand the 1-form symmetry through the quasiparticles of this theory, which are described by the line operators 
\begin{equation}
    \mathcal{W}_{C_1, \xi } = e^{i \int_{C_1} \xi_i b^i }\,, \qquad \xi \in \ZZ^n\,,
\end{equation}
where $C_1$ is a 1-cycle. These operators have braiding, spin (in the sense of self-statistics) and charge
\begin{equation} \label{eq:braidspincharge}
    B(\xi,\zeta) = e^{2\pi i \xi K^{-1} \zeta}\,, \qquad S(\xi) = \frac{1}{2}\xi^T K^{-1} \xi\,, \qquad Q(\xi) =  \xi^T K^{-1} t\,,
\end{equation}
which are only defined modulo 1. Quasiparticles with non-trivial braiding are anyons, while the transparent operators, those with trivial braiding, correspond to the worldlines of particles. The operators with $\xi = K \zeta$ for $\zeta \in \ZZ^n$ braid trivially with all other lines, and so are transparent. The set of non-transparent lines live in $G_{\text{one-form}} = \faktor{\ZZ^n}{K\ZZ^n} $. This group is finitely generated, with the different generators commuting with each other. 

This one-form symmetry has an 't Hooft anomaly if the 1-form symmetry cannot be consistently gauged, or equivalently if the non-transparent lines have non-integer spin \cite{Hsin_2019}. If $G_{\text{one-form}} =\ZZ_{N}$ then the specific t' Hooft anomaly is characterized by the integer $p \mod N$ such that the generator of the $\ZZ_N$ 1-form symmetry has spin $\frac{p}{2N}$.\footnote{Note that $p$ is defined modulo $N$ as these are spin theories, in which case we can multiply the generator by the transparent fermion to get a new generator with spin shifted by $1/2$. If these were bosonic theories with no transparent fermion then $p$ is defined modulo $2N$ and must be even \cite{Hsin_2019}.} If $G_{\text{one-form}}$ is finitely generated by several lines as in \eqref{eq:g1form}, then the t' Hooft anomaly will be characterized by the set of such integers, one for each generator.

One prominent choice of $K$ matrices, and those that appear throughout our construction, are of the form
\begin{equation}
    K = \begin{pmatrix}
        m_1 & \pm 1 & 0 & 0 & \cdots \\
        \pm1 & m_2 & \pm1 & 0 & \cdots\\
        0 & \pm1 & m_3 & \pm1  & \cdots\\
        0 & 0 & \pm1 & m_4 & \ddots \\
        \vdots & \vdots & \vdots & \ddots & \ddots \,.
    \end{pmatrix}
\end{equation}
 The Smith normal form of this $K$ is the diagonal matrix $B = \text{Diag}(1, \ldots, 1, \det(K) )$. This can be shown by using the fact that the diagonal elements in the Smith normal form are $b_j = d_j/d_{j-1}$, where $d_j$ is the greatest common divisor of the determinants of the $j\times j$ minors of $K$. Since $K$ is an $n\times n$ matrix, then for each $j<n$ there is a triangular minor of size $j\times j$ with $\pm1$'s on the diagonal, so it follows that $d_j = 1$ for all $j<n$, proving the form of $B$. Thus the one-form symmetry group for theories with these $K$ matrices is simply $G_{\text{one-form}} = \ZZ_{|\det(K)|}$.

These results persist when some of the dynamical fields are spin$_c$ fields, with only a minor modification to the spin of quasiparticles. To understand why, consider a more general theory with both $U(1)$ and spin$_c$ fields
\begin{equation}
    \mathcal{L} =  -\frac{K_{ij}^{b}}{4\pi}  b^i  d b^j - k_\alpha^a CS(a^\alpha,g) -\frac{K_{j\alpha}^{ab}}{4\pi}  b^j  d a^\alpha
    + \frac{t_j^b}{2\pi} b^j dA, 
\end{equation}
where $\alpha\in \ZZ_m$, $K^b\in \ZZ^{n\times n} $ is a symmetric matrix, and $K^{ab}\in \ZZ^{m\times n}$. The spin-charge relation mandates $K_{jj}^b + t_j + \sum_\alpha K_{j\alpha}^{ab} \in 2\ZZ$. One can then build a $K$ matrix and $t$ vector for the larger theory as $K = \begin{pmatrix} K^b & K^{ab} \\ {K^{ab}}^T & K^a \end{pmatrix}$, $t = \begin{pmatrix} t^b \\ 0_m\end{pmatrix}$ where $K^a$ is a diagonal matrix with $k^a_\alpha$ on the diagonal. Line operators still correspond to Wilson loops
\begin{equation}
    \mathcal{W}_{C_1, \xi } = e^{i \int_{C_1} \left(\xi_i^b b^i + \xi_\alpha^a\left( a^\alpha + \frac{\omega}{2} \right)\right) }\,, \qquad \xi \in \ZZ^{n+m}\,,
\end{equation}
and their braiding and charge are identical to those in \eqref{eq:braidspincharge}. The spin, however, is modified to  
\begin{equation}
    S(\xi) = \frac{1}{2}\xi^T K^{-1} \xi\ + \sum_{\alpha = 1}^m\frac{\xi^a_\alpha}{2}.
\end{equation}
As the braiding of these lines is exactly identical to the previous analysis, the non-transparent lines still live in $G_{\text{one-form}} = \faktor{\ZZ^{n+m}}{K\ZZ^{n+m} }$, and so the 1-form symmetry group is the same. Thus, the Abelian state we construct at filling $p/q$ with $q$ odd have a 1-form symmetry $G_{\text{one-form}} = \ZZ_{|\det(K)|} = \ZZ_q$. Furthermore, it is simple to check that the line with $\xi = t$ generates the 1-form symmetry, so the t' Hooft anomaly of this $\ZZ_q$ 1-form symmetry is $p$. In fact this tells us that these are all just the minimal Abelian TQFTs $\mathcal{A}^{q,p}$ described in \cite{Hsin_2019}.

\bibliographystyle{JHEP}
\bibliography{refs}

\end{document}